\documentclass[superscriptaddress,aps,amsmath,amssymb,showpacs,showkeys]{revtex4-2}
\usepackage{natbib} 
\usepackage[dvips]{graphicx}
\usepackage{times}
\usepackage{braket}
\usepackage{xcolor}
\usepackage{orcidlink}
\usepackage{hyperref}
\usepackage{booktabs}
\usepackage{float} 
\usepackage{subfigure}
\usepackage{siunitx}
\hypersetup{
	colorlinks=true,
	urlcolor=magenta,
	linkcolor=red,
	citecolor=blue
}

\begin{document}
\title{Mapping Nearby Galaxies with Apache Point Observatory: Group and field 
galaxies' morphologies in the colour-magnitude plane}
	
\author{Pius Privatus\orcidlink{0000-0002-6981-717X}}
\email[Email: ]{privatuspius08@gmail.com}
\affiliation{Department of Physics, Dibrugarh University, Dibrugarh 786004, Assam, India}
\affiliation{Department of Natural Sciences, Mbeya University of Science and Technology, Iyunga 53119, Mbeya, Tanzania}
	
\author{Umananda Dev Goswami\orcidlink{0000-0003-0012-7549}}
\email[Email: ]{umananda@dibru.ac.in}
\affiliation{Department of Physics, Dibrugarh University, Dibrugarh 786004, Assam, India}
	
\begin{abstract}
This study involves the use of integral field spectroscopy (IFS) data from 
Mapping Nearby Galaxies at Apache Point Observatory (MaNGA) to investigate 
whether the morphology influences the environmental dependence of galaxies' 
colours and the colour-magnitude planes. The galaxies are classified into 
six morphologies (Elliptical, Lenticular, Early-type, Intermediate-type,  
Late-type spirals and Irregular) and further in field and group environments. 
The distributions of colours ($B-V, B-R, u-g, g-r, r-i$ and $i-z$) are 
compared between field and group environments and then the colour-magnitude 
planes are analysed. It is observed that Intermediate and Late-types spirals 
preferentially exist in field environments while Early-type spirals exist in 
groups. The colours and colour-magnitude planes of Elliptical, Lenticular, 
Early-type and Intermediate-type spirals depend on the environment while for 
the Late-type and Irregular galaxies, their dependence on the environment is 
very weak. The study concludes that the dependence of colours and 
colour-magnitude planes on the environment is influenced by morphology.
\end{abstract}
\maketitle    

\section{Introduction}
Morphology and colour are two among the fundamental properties in galaxy 
evolution, with which galaxies are classified into late-type (with 
disk-dominated, spiral arms and young stellar populations) in blue colour 
and early-type (with bulge-dominated, elliptical and old stellar populations) 
in red colour \cite{strateva2001color, correa2017relation,dullo2020black, bom2024extended,mendes2019southern}. 
Ref.\ \cite{strateva2001color} observed the existence of a strong correlation 
between colour and morphology, due to which elliptical and lenticular tend to 
be redder than spiral and irregular galaxies. The findings by 
Ref.\ \cite{bom2024extended} pointed out that late-type display blue colour 
than early-type galaxies, using a Deep Learning (DL) based morphological 
catalogue built from images obtained by the third release of Southern 
Photometric Local Universe Survey (S-PLUS) \cite{mendes2019southern}. 
The morphological changes 
are usually accompanied by colour transformation, however it should be kept in 
mind that it is not necessarily a case of vice versa as spiral galaxies may 
transform from blue cloud to red sequence without changing from 
late-type to early-type \cite{liu2019morphological,correa2019origin}. 
Furthermore, the study by 
Ref.\ \cite{smethurst2022quantifying} observed that the use of colour alone 
as a reliable indicator of galaxy morphology, leads to significant 
contamination and misclassification, hence it is not valid to use colour as 
the only factor for galaxies' morphological classification.
   
A number of studies have shown galaxy morphology to depend on the environment 
using morphology-density relation, with early-type galaxies located in dense, 
while late-type galaxies in low-density environments \cite{mei2023morphology, sazonova2020morphology,ball2008galaxy,paulino2019vis3cos,
gu2021effect,postman1984morphology,
goto2003morphology}. The study by Ref.\ \cite{mei2023morphology} observed that 
the galaxy morphology to be highly influenced by the environment pointing out 
quenching at which the star formation of galaxies decreases and mergers to be 
significantly dense when compared to low-density regions which are 
responsible for morphological changes \cite{sazonova2020morphology}. 
Ref.\ \cite{paulino2019vis3cos}, studying the impact of local density on 
the morphology for $\sim 500$ star forming and quiescent galaxies concluded 
that the process affecting a galaxy's star formation must also affect the 
morphology. Similarly, Ref.\ \cite{gu2021effect}, exploring the influence of 
environment on star formation and morphology using the sample from the Cosmic 
Assembly Near-infrared Deep Extragalactic Legacy Survey (CANDELS) as detailed 
in Refs.\ \cite{grogin2011candels,koekemoer2011candels}, within the redshift of
range $0.5<z<2.5$ observed that the morphology is connected to galaxy's star 
formation. 
   
However, the study by Ref.\ \cite{cooke2023roles}, investigating the 
relationship between the morphology, star formation rate (SFR) and environment 
using the sample from the Cosmic Evolution Survey (COSMOS) as detailed in 
Ref.\ \cite{scoville2007cosmic}, found that the shape of the main sequence in 
the relation of colour against colour and colour against  specific star 
formation rate (SSFR) including the turnover at higher stellar mass does not 
depend on the environment for the redshift range $0< z< 3.5$. 
Ref.\ \cite{perez2023relation} mention the environment to play a 
significant role in shaping the morphology of galaxies, so should be taken 
into account to pinpoint the mechanism driving the influence of clusters in 
galaxy evolution further insisted that the morphology should be taken into 
account when discussing the influence of the environment on galaxy evolution. 
Using a limited sample in volume and stellar mass from the sixteen release of 
the Sloan Digital Sky Survey (SDSS) Ref.\ \cite{bhattacharjee2020can} obtained 
that the existing relations of morphology-stellar mass and again 
environment-stellar mass are not adequate in explaining the relation between  
morphology and environment since the factors responsible to alter morphology 
are not always necessary to have a direct effect on stellar mass and the vice 
versa, despite of the widely accepted possibility of a relation between 
morphology and environment to be linked with stellar mass.
   
On the other hand, a number of studies using single-fibre spectroscopy have 
shown galaxy colours to depend on the environment, with blue galaxies found 
in the low-density environment while red is located in a dense environment 
\cite{zhang2015ur,blanton2005relationship,brown2000clustering,
zehavi2005luminosity,weinmann2006properties,deng2014environmental, 
bamford2008revealing,skibba2009halo,tinker2008void,pandey2020exploring}. 
Ref.\ \cite{brown2000clustering} used the colour selection to study the 
clustering of galaxies within the redshift range $0<z<0.4$, obtained that 
clustering strongly depends on the environment with blue galaxies weakly 
clustered than the red ones suggesting that galaxy clustering strongly
correlate with colour than morphology. Ref.\ \cite{blanton2005relationship} 
using the data from SDSS, concluded that colour is the most predictive 
physical property of the local environment. The results of this reference 
suggest that galaxy structural properties are less correlated with the 
galaxy's environment than the stellar mass (M$\star$) and SFR. In the study on 
the galaxy colours in different environments obtained 
by considering the galaxy's local dimension using a volume limited sample 
constructed from SDSS, Ref.\ \cite{pandey2020exploring} obtained that the red 
galaxies are preferentially found in filament with high fraction when 
compared to sheet environment no matter the luminosity of a particular galaxy. 
Making the local density fixed authors observed the same trend meaning that 
the colours of galaxies depend on the environment even if the local density 
is fixed. Furthermore, it was found that the fraction of red galaxies 
increases with the local density. From this, they concluded the existence of 
colour bimodality in all environments and luminosities implying that the blue 
cloud to red sequence transformation exists in all environments. Similarly, 
Ref.\ \cite{zhang2015ur} concluded that blue galaxies dominate the isolated 
and small groups, while red galaxies dominate dense groups and clusters.
   
The study by Ref.\ \cite{skibba2009galaxy}, using the data from the galaxy zoo 
project as detailed in Ref.\ \cite{lintott2011galaxy}, found that a galaxy's 
colour depends on the environment at a fixed morphology although they further 
obtained that the relation between morphology and environment is very weak 
when colour is kept fixed. Analysing the correlation between galaxy 
morphology, colour, stellar mass and environment using the density approach, 
Ref.\ \cite{bamford2009galaxy} observed that morphology and colour are 
different functions of environment and both depend on stellar mass as
once the stellar mass is fixed morphology weakly depends
on the environment while galaxy colour strongly depends on the environment.
They pointed out that galaxies with higher stellar mass are redder than the 
ones with lower stellar mass irrespective of their morphology in all 
environments. The galaxies having low M$\star$ are most cases bluer 
and found in low-density then become redder in high-density environments 
again irrespective of their morphology. In this sense they
concluded that galaxy colour strongly depends 
on the environment rather than the morphology. Studying nearby massive galaxies 
specifically on the existence of a relation between the star formation, 
morphology and environment, Ref.\ \cite{bait2017interdependence} observed the 
existence of a correlation between morphology and SSFR. On the other hand, 
authors observed a weak dependence of morphology and SFR on 
the environment. The study concluded that the responsible physical process 
that shapes galaxy morphology also determines the rate of forming stars in a 
particular galaxy. Ref.\ \cite{dhiwar2023witnessing}, studying the evolution 
of Luminosity ($L^*$) of elliptical galaxies using a sample of $36500$ galaxies 
from the SDSS where $51$ ellipticals were selected with $12$ blue cloud, 
$11$ green valley and $28$ red sequence ellipticals, observed that galaxy 
colour is related to star formation keeping the morphology constant as they 
concluded that most of the red sequence and green valley ellipticals are still 
forming stars or have been quenched recently while the blue cloud ellipticals 
are vigorously forming stars, which is not concentrated only at the centre 
but extend over the entire dimensions of a particular galaxy.
   
Using a sample of $49911$ galaxies obtained from Galaxy And Mass Assembly 
survey (GAMA) as detailed in Refs.\ \cite{driver2022galaxy,baldry2018galaxy,
driver2011galaxy,driver2009gama} with a redshift range between $0.05$ and 
$0.18$, to investigate how red galaxies fraction varies with environmental 
measurements, Ref.\ \cite{bhambhani2023red} found that the fraction varies 
with environment, however different density measurement results to variation 
in information. Analysing the galaxy's properties and their dependence on the 
environment for luminous and non-luminous (separated by a characteristic 
magnitude M$^*\sim 19.7$ mag), Ref.\ \cite{deng2009comparisons} compared the 
properties in lowest and highest density regime using the radius to the fifth 
nearest neighbour to characterize galaxy environment. Authors observed 
luminosity-environment variations as a strong dependence on the environment 
was observed for luminous galaxies having luminosity above the characteristic 
magnitude while for the case of non-luminous galaxies that are having 
luminosity below the characteristic magnitude, a weak dependence on the 
environment was observed. Authors further concluded that colour, morphology 
and concentration index (ci) strongly depend on the environment no matter the
value of luminosity. This implies that the characteristic magnitude is the 
parameter guiding only the dependence of luminosity on the environment.
    
Ref.\ \cite{bernardi2003early}, studying colours of early-type galaxies 
selected from SDSS at $z<0.3$ observed that when a relation connecting
the galaxy's effective radius, 
mean surface brightness, 
and central velocity dispersion (the fundamental plane) is 
considered, the galaxies in a less dense environment are not significantly 
different from dense environment. Furthermore, the colour magnitude diagram 
of a dense region is not statistically different from that of less dense 
regions which implies that colour magnitude diagram is not influenced by 
density. The study by Ref.\ \cite{deng2013environmental} found that the 
galaxy colours depend on the environment using a sample of galaxies from 
SDSS opposite to Ref.\ \cite{deng2023environmental}, which used a sample of 
galaxies from SDSS also to study the dependence colour of active 
galactic nuclei (AGNs), concluded that all five colours ($u-r, u-g, g-r, r-i$ 
and $ i-z$) of AGNs weakly depend on the environment. On other side the 
colours of galaxies are related with the cosmic-web environment where the 
green-valley ellipticals are found in either of the environment, the blue cloud 
preferring low density region while the red sequence ellipticals reside in 
dense environment \cite{o2024effect}. The existence of contradictory 
results draws attention to studying these relations in a concise manner using 
different samples and approaches.
   
In this study we aim to use integral field spectroscopy (IFS) data from the 
Mapping Nearby Galaxies survey at Apache Point Observatory (MaNGA) Ref.\ \cite{bundy2014overview,blanton2017sloan} 
to investigate if morphology influences the dependence of galaxy colours and 
the colour-magnitude plane on the environment. The layout of this paper is as 
follows. We present the method of getting data in Section \ref{secII}. Our 
findings are presented in Section \ref{secIII} and in Section \ref{secIV} we 
discuss the findings. The summary and conclusion of the study are presented 
in Section \ref{secV}. Throughout this paper we adopt a standard cosmology 
with the Hubble constant $H_0=63$ km s$^{-1}$ Mpc$^{-1}$, density of matter 
$\Omega_{m}=0.3$, and dark energy density $\Omega_{\Lambda}=0.7$.
   
\section{Data} 
\label{secII}
\subsection{MaNGA survey}\label{mg}
The Mapping Nearby Galaxies at Apache Point Observatory (MaNGA) survey Ref.\ \cite{blanton2017sloan} 
is one of the currently available integral-field spectroscopy (IFS) surveys 
containing the largest sample of galaxies employing fiber-bundle-based 
integral field units (IFUs). The spatially resolved spectroscopic measurements 
contains $\sim 10000$ galaxies, employing seventeen fiber-IFU operating in 
the range from $12$ to $32$ 
arcseconds ($19-127$ fiber per IFU), covering the wavelength range of 
$3600-10300 \AA$ at a resolution of $R \sim 2000$ \cite{bundy2014overview,
blanton2017sloan,drory2015manga}. The target for the MaNGA survey was chosen 
considering the mass ranges and galaxy colours whereby the sample is laying 
within the redshift (z) between $0.01$ to $ 0.15$. The survey involves three 
subsamples where the primary sample has optimised spatial coverage out 
to $\sim1.5$ of effective radius of galaxies (Re) making $\sim 50\%$ of the 
total sample. The secondary sample extends the radial coverage to $\sim 2.5$ Re 
making $\sim 33\%$ of the total sample and the colour-enhanced subsample 
making $\sim 17\%$ of total \cite{wake2017sdss,yan2016sdss}. It is very 
important to remember the statistical effects produced by the use of 
flux-limited samples where at higher redshift the brighter galaxies obscure 
the fainter one (Malmquist bias) \cite{teerikorpi2015eddington}. Thus 
to mitigate the Malmquist bias we use the volume-limited samples with 
redshift (z) not exceeding $0.15$, minimum angular separation to avoid fiber 
collisions ($\theta_{\text{min}} = 55$ arcsecond), with 
$500$ km/s line-of-sight velocity difference ($\Delta V$) up to $1$ Mpc 
projected distance ($R_p$) \cite{abdurro2022seventeenth,
argudo2015catalogues}. The volume-limited sample was selected using 
a selection function given by
\begin{equation}
	S(z, \theta, R_p, \Delta V) =
	\begin{cases}
				1, & \begin{aligned}
			&0.01 \leq z \leq 0.15,\quad \theta \geq \theta_{\text{min}},
			\quad R_p \leq 1\ \text{Mpc},\quad \Delta V \leq 500\ \text{km/s},
		\end{aligned} \\
		0, & \text{otherwise}.
	\end{cases}
	\label{SP}
\end{equation}

\subsection{Galaxies morphologies}\label{gm}
\begin{figure}[h!]
	\subfigure{
		\includegraphics[width=0.3\linewidth]{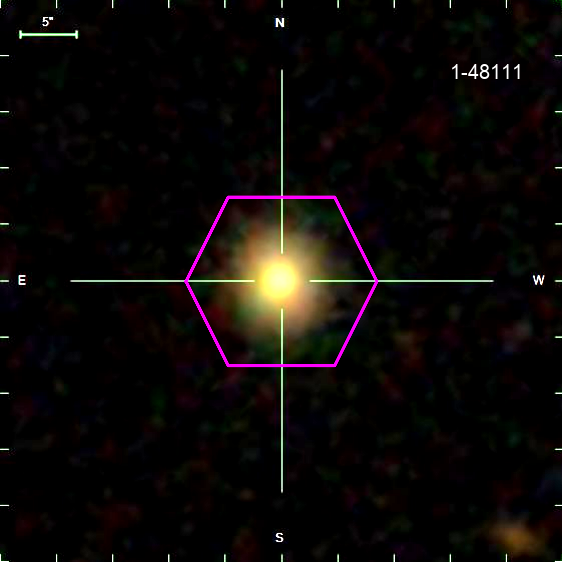}
	}
	\hspace{0.1cm}
	\subfigure{
		\includegraphics[width=0.3\linewidth]{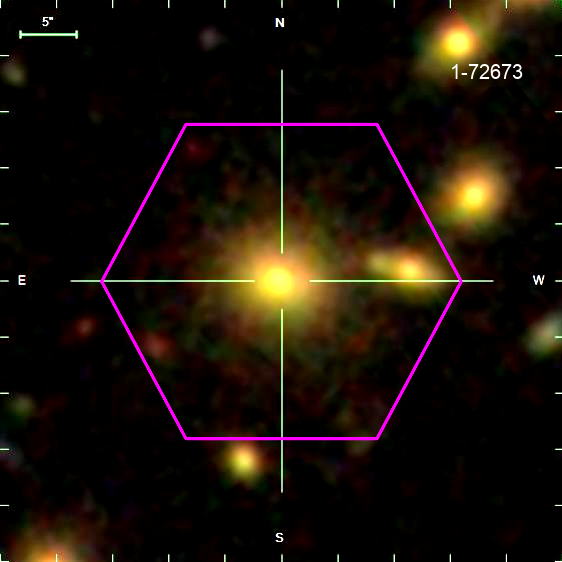}
	}
	\hspace{0.1cm}
	\subfigure{
		\includegraphics[width=0.3\linewidth]{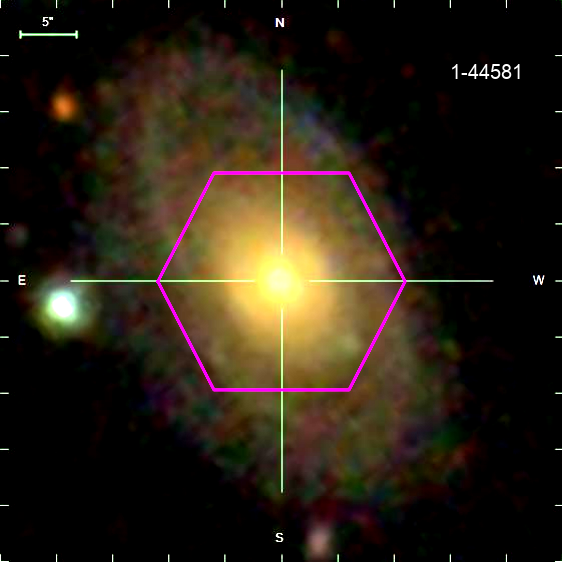}
	}
	\hspace{0.1cm}
	\subfigure{
		\includegraphics[width=0.3\linewidth]{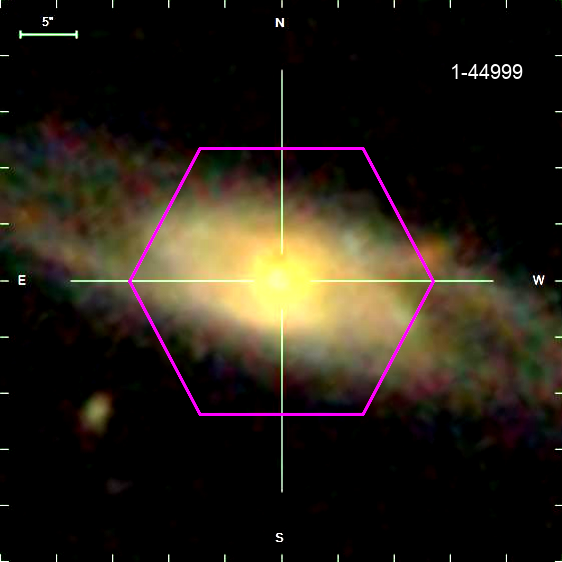}
	}
	\hspace{0.1cm}
	\subfigure{
		\includegraphics[width=0.3\linewidth]{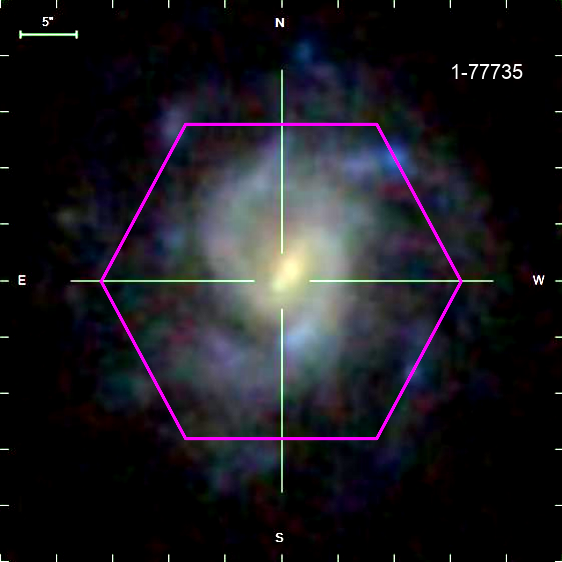}
	}
	\hspace{0.1cm}
	\subfigure{
		\includegraphics[width=0.3\linewidth]{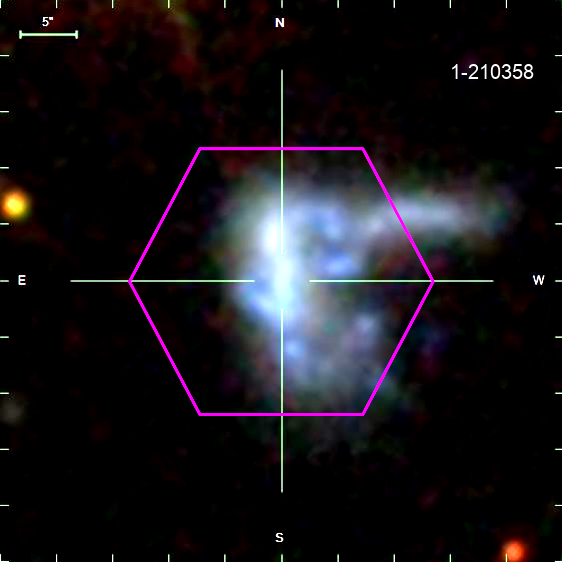}
	}
	\vspace{-0.2cm}
	\caption{The images for six classified morphologies: Elliptical (top left), 
		Lenticular (top middle), Early-type spiral (top right),  Intermediate-type 
		spiral (bottom left), Late-type spiral (bottom middle) and 
		Irregular (bottom right). The top right white text in each image indicates 
		the mangaid while the pink hexagon covers the spatial extent of the MaNGA 
		Integral Field Unit.}
	\label{IM}
\end{figure}

In this study we use the accurate visual morphology as detailed in 
Ref.\ \cite{vazquez2022sdss}, which follows the basic Hubble classification 
scheme \cite{hubble1926}. The probability of a galaxy falling in one of the
classes of the classification scheme for $9652$ galaxies obtained from 
subsection \ref{mg} are given in 
Ref.\ \cite{sanchez2022sdss}. The best morphological type 
(best$_{-}$type$_{-}$n) is given by integers ($-2, -1, 0, 1, 2, 3, 4, 5, 6, 
7, 8, 9$ and $10$) arranged from early-type to late-type morphologies that are 
central dominant galaxies which are massive elliptical (cD) to Irregular (Irr). 
We classified the galaxies into six morphologies: cD  and E (Elliptical), 
SO (Lenticular), Sa, Sab and Sb (Early-type spirals), Sbc and Sc 
(Intermediate-type spirals), Scd and Sd (Late-type spirals), Sdm, Sm and Irr 
(Irregular). A total number of $926\,(9.59\%)$, $1817\,(18.83\%)$, 
$3798\,(39.35\%)$, $2838\, (29.40\%)$, $185\,(1.92\%)$, $88\,(0.91\%)$ for 
Elliptical (EL), Lenticular (LE), Early-type spirals (ES), Intermediate-type 
spirals (IS), Late-type spirals (LS) and 
Irregular galaxies (IR), respectively were obtained. The sample images for 
each of the six classified morphologies 
used in this study are shown in Fig.~\ref{IM}.

\subsection{Galaxy environment}\label{en}
\begin{figure}[h!]
	\subfigure{
		\includegraphics[width=0.31\linewidth]{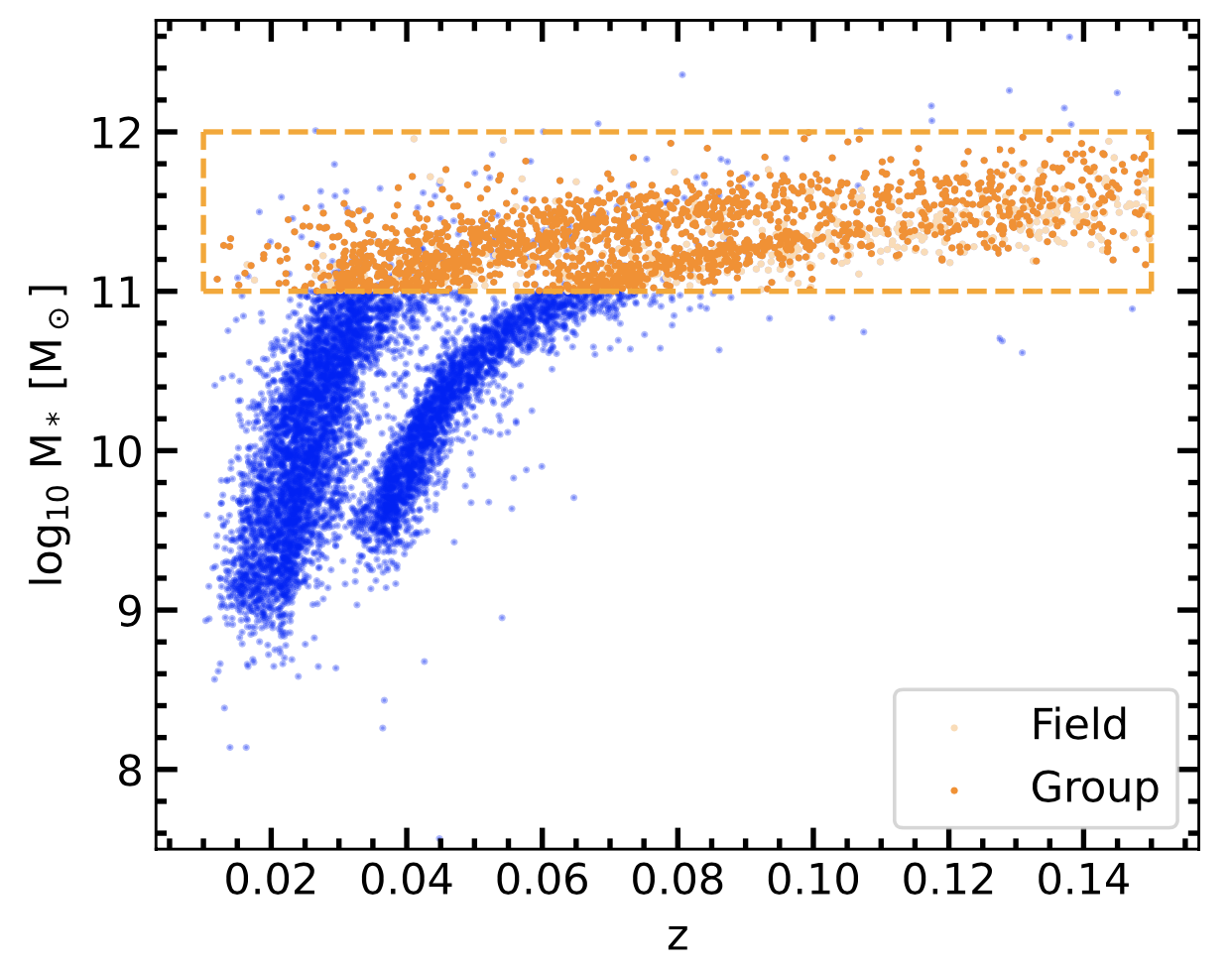}
	}
	\subfigure{
		\includegraphics[width=0.32\linewidth]{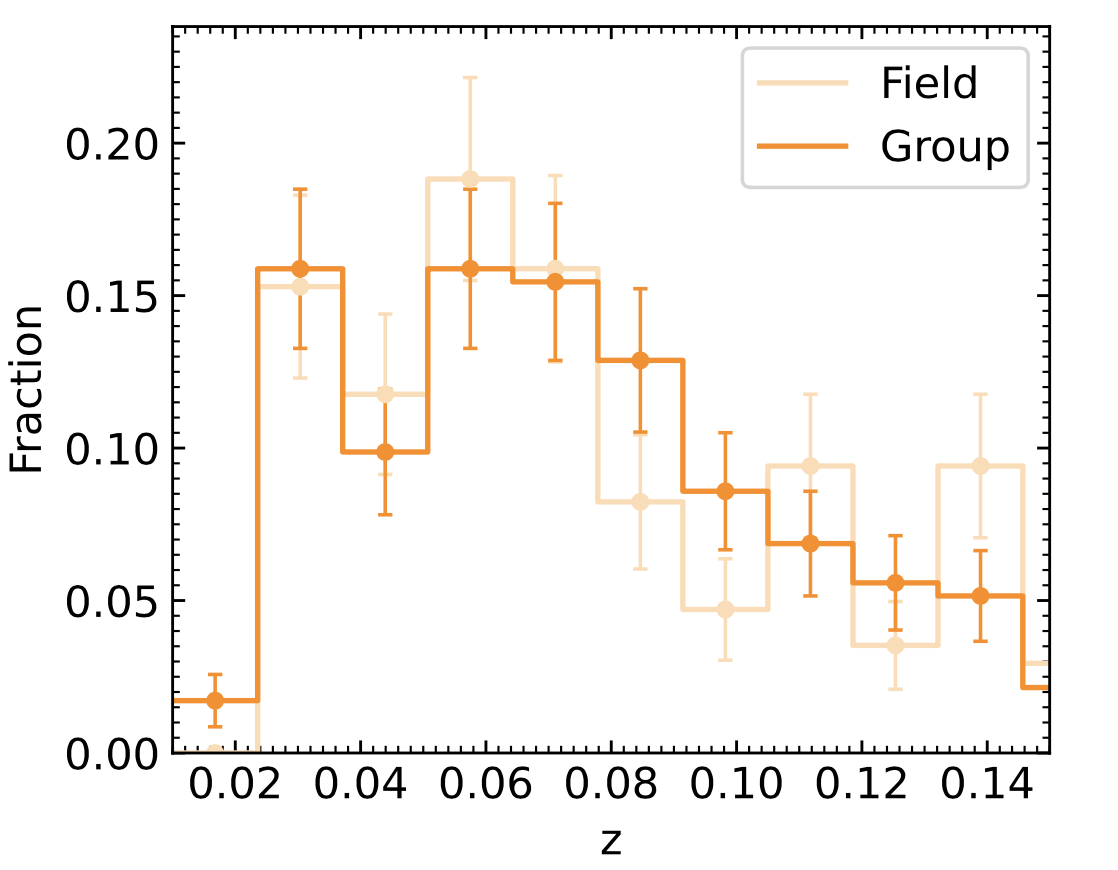}
	}
	\subfigure{
		\includegraphics[width=0.32\linewidth]{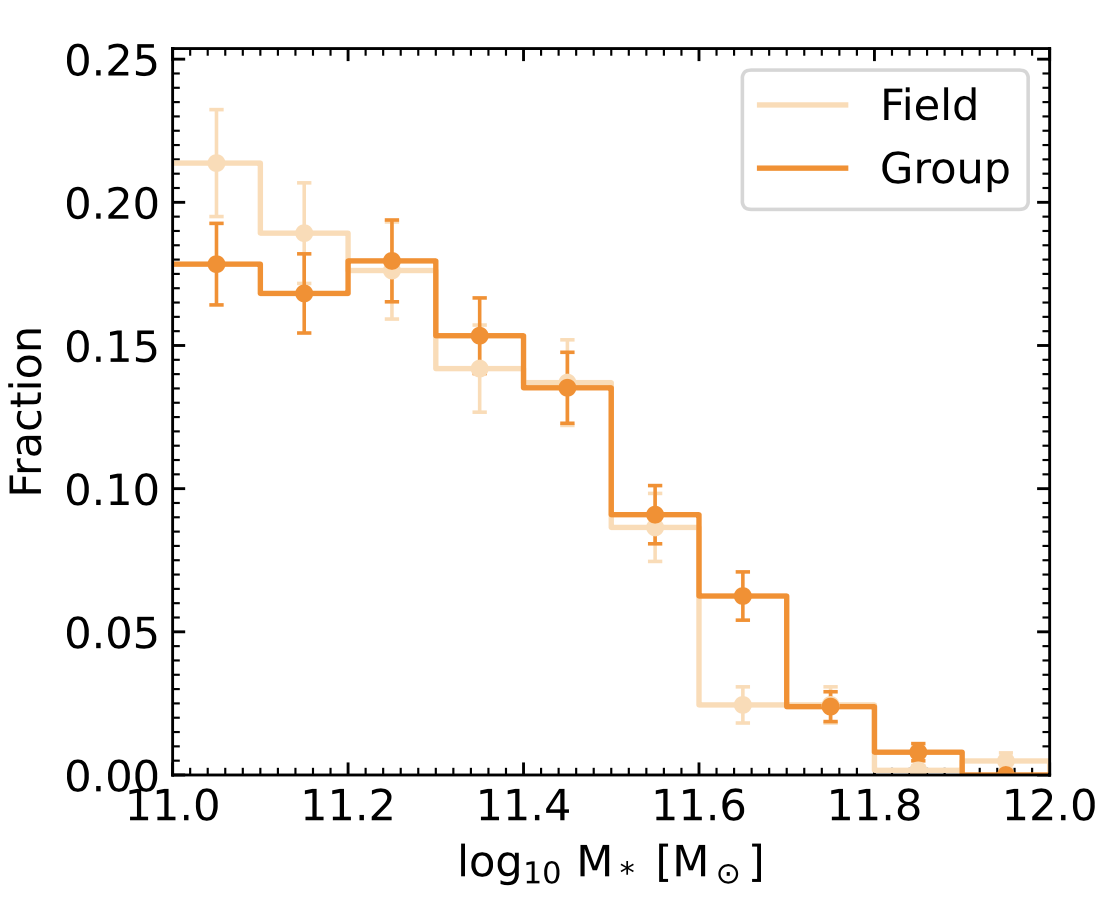}
	}
	\hspace{-0.2cm}
	\caption{Redshift- stellar mass scatter plot (left panel), 
		redshift (middle panel), stellar mass (right panel) distributions for field (light) 
		and group (dark) galaxies sample. 
		The error bars presented in this figure and other similar 
		figures are for $1\sigma$ Poissonian errors.}
	\label{HL}
\end{figure}

The galaxy environments were quantified using the Galaxy Environment for MaNGA 
Value Added Catalogue (GEMA-VAC) as detailed in 
Refs.\ \cite{etherington2015measuring,wang2016elucid}. 
Based on the information provided in GEMA-VAC, the galaxies are assigned in 
groups by means of a halo-based group finder as detailed in 
Ref.\ \cite{yang2007galaxy}. Some galaxies have more than one neighbour 
(Group size (GS) $\geq2$) while others have no nearby galaxy (GS $=1$). 
Throughout this study the galaxies have stellar mass ranges 
$11\leq log_{10}$ M$\star \leq12$ (M$\odot$), the one with GS $\geq2$ criteria 
are considered as group (G) galaxies, while the ones with GS $= 1$ criteria 
are taken as field (F) galaxies. The panels of Fig.~\ref{HL}
show the volume limited samples illustrated using the 
redshift-stellar mass diagram (left panel), the redshift (middle panel) and 
stellar mass (right panel) distributions for field and group galaxies. We 
performed the Kolmogorov-Smirnov (KS) test \cite{hodges1958significance,
harari2009kolmogorov} and Anderson–Darling (AD) \cite{anderson1952asymptotic,
pettitt1976two,scholz1987k,babu2006astrostatistics} statistical test, keeping 
in mind that the lower (close to zero) KS, AD statistics values and the 
higher p-values ($\geq0.05$) indicate that the two distributions are similar, 
while the higher KS, AD statistics values and the lower p-values ($<0.05$) 
indicate that the two distributions are different 
\cite{harari2009kolmogorov,scholz1987k,babu2006astrostatistics}.  
The volume limited samples have KS statistics (p-value) of $0.06$ 
($0.814$),  $0.07$ ($0.074$) and AD statistics (p-value) of $0.05$ ($0.25$), 
$2.62$ ($0.274$) for redshift and stellar mass,  respectively. This indicates 
that there is no significant difference between field, group redshift and 
stellar masses distributions, supporting the observations from histograms in 
Fig.~\ref{HL}.

Using the stated conditions the following 
numbers of galaxies were obtained. For Elliptical: $247\,(26.67\%)$ and 
$679\,(73.33\%)$, for Lenticular: $558\,(30.71\%)$ and $1259\,(69.29\%)$, 
for Early-type spiral: 
$1577\,(41.52\%)$ and $2221\,(58.48\%)$, for Intermediate-type spiral: 
$1685\,(59.37\%)$ and $1153\,(40.63\%)$, for Late-type spiral: 
$110\,(59.46\%)$ and $75\,(40.54\%)$, for Irregular: $38\,(43.18\%)$ and 
$50\,(56.82\%)$ as field and group galaxies, respectively. The classified 
subsamples were used for the analysis in the next sections unless otherwise 
stated.
\section{Results}\label{secIII}
\begin{figure}[h!]
  	\subfigure{
  		\includegraphics[width=0.3\linewidth]{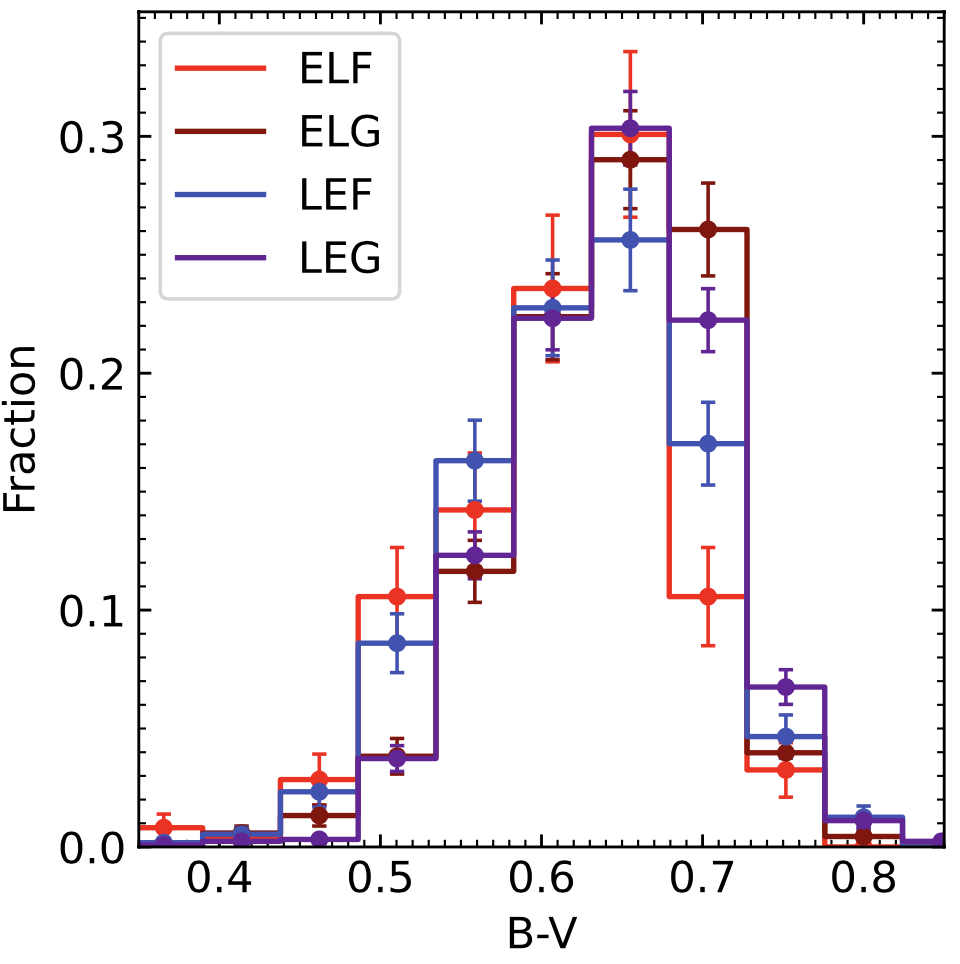}
  		}
  		\hspace{0.3cm}
  		\subfigure{
  		\includegraphics[width=0.3\linewidth]{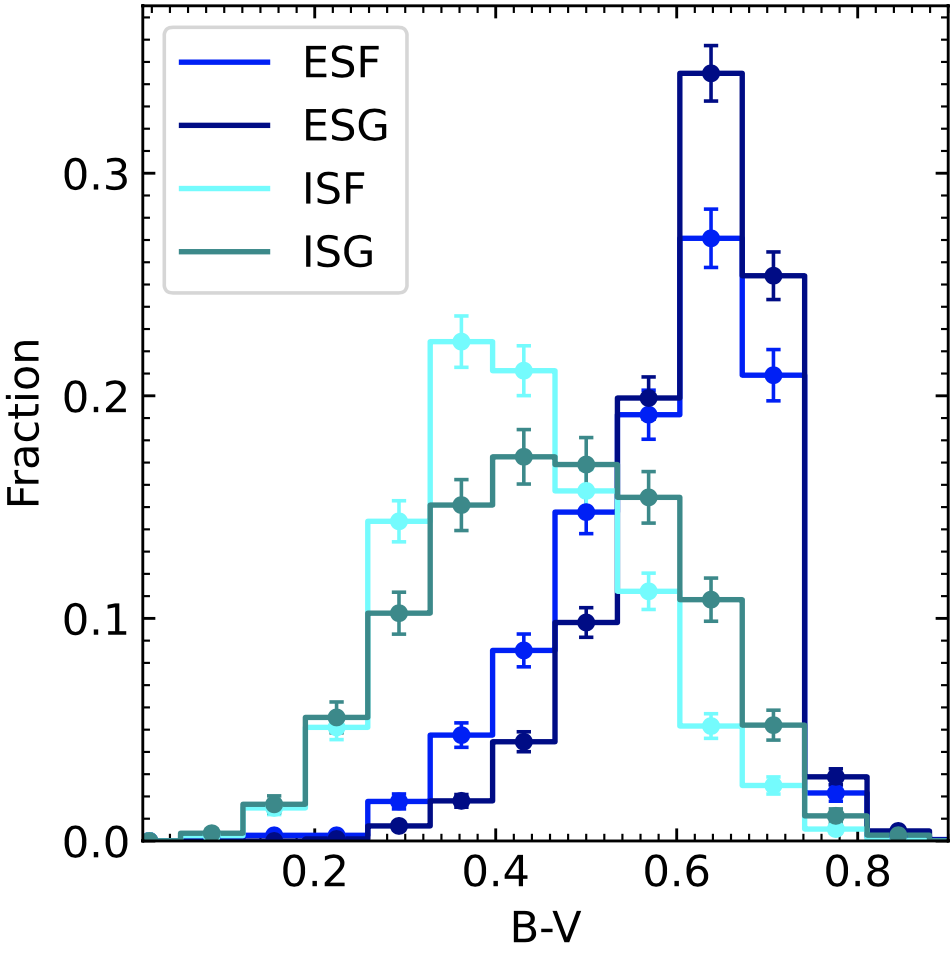}
  		}
  		\hspace{0.3cm}
  		\subfigure{
  		\includegraphics[width=0.3\linewidth]{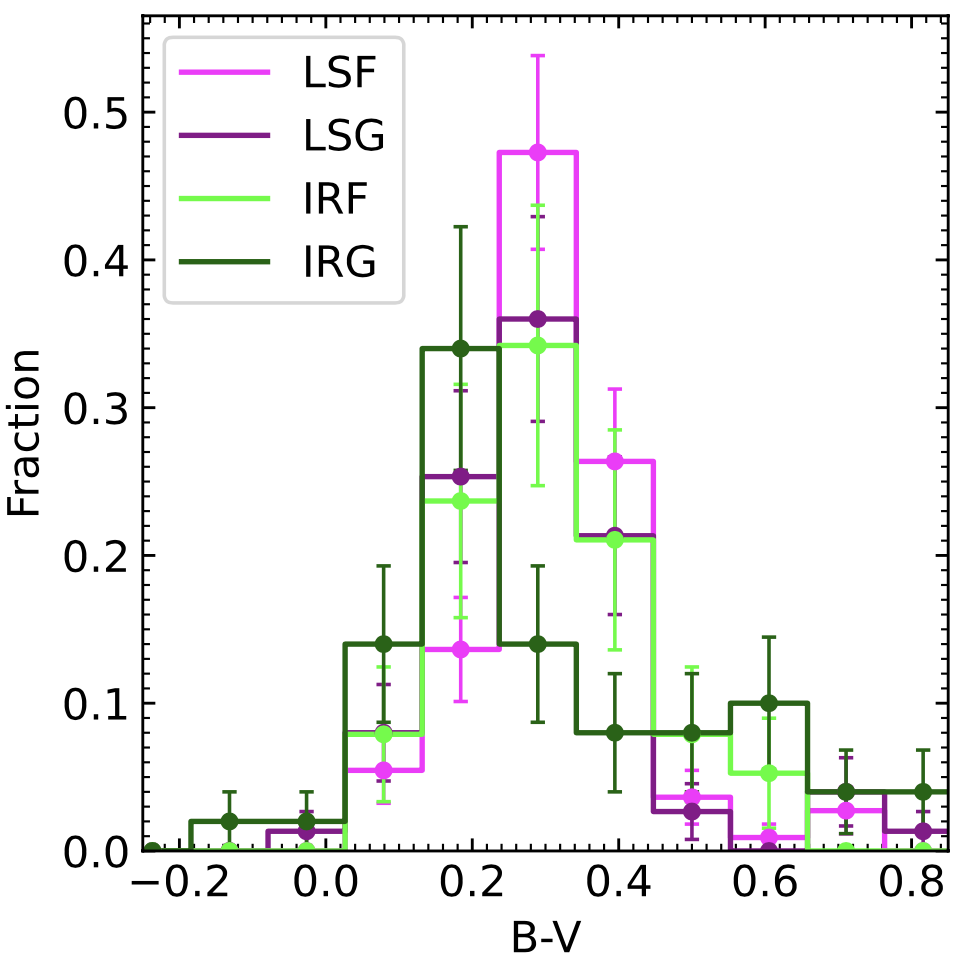}
  		}
        	\vspace{-0.2cm}
\caption{Comparison between field (light colour) and group (dark 
colour) galaxies' $B-V$ 
colour distributions. Left panel: Elliptical (red) and Lenticular (indigo); 
Middle panel: Early-type spiral (blue) and Intermediate-type spiral (cyan); 
Right panel: Late-type spiral (magenta) and Irregular (green) galaxies.}
\label{BV}
\end{figure}

\begin{figure}[h!]
  		\subfigure{
  			\includegraphics[width=0.3\linewidth]{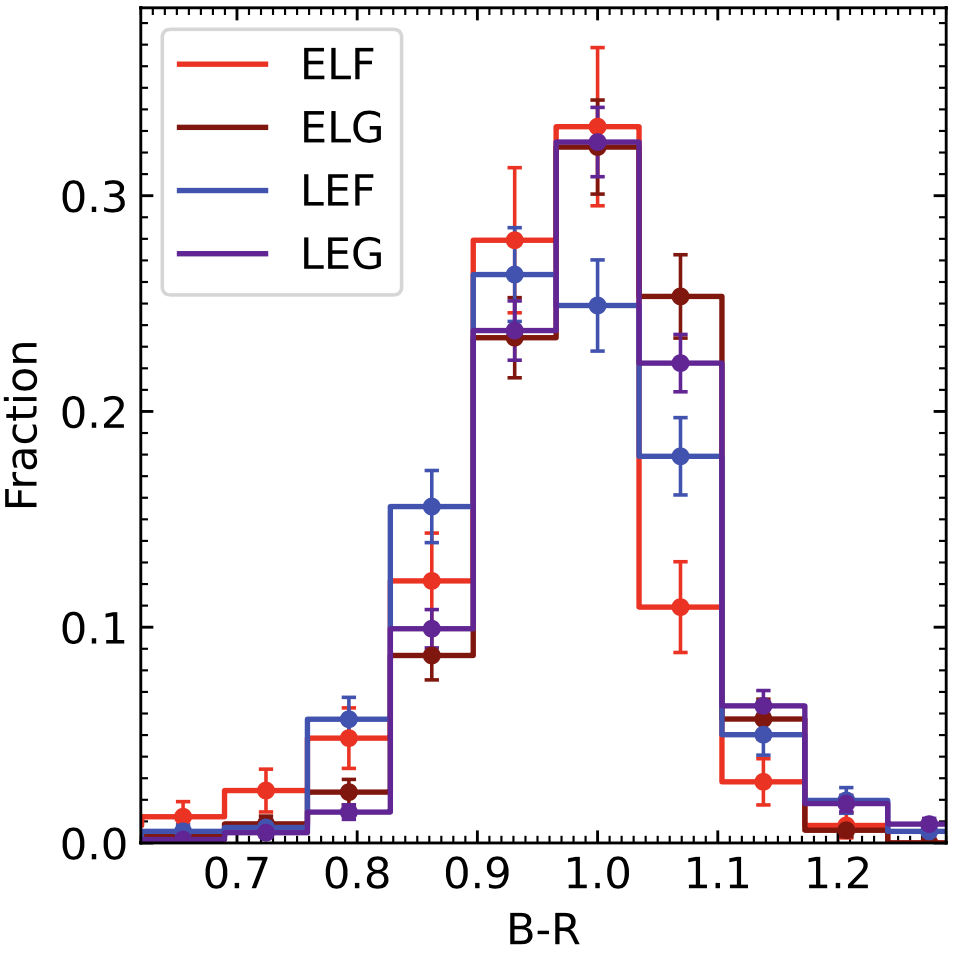}
  		}
  			\hspace{0.3cm}
  		\subfigure{
  			\includegraphics[width=0.3\linewidth]{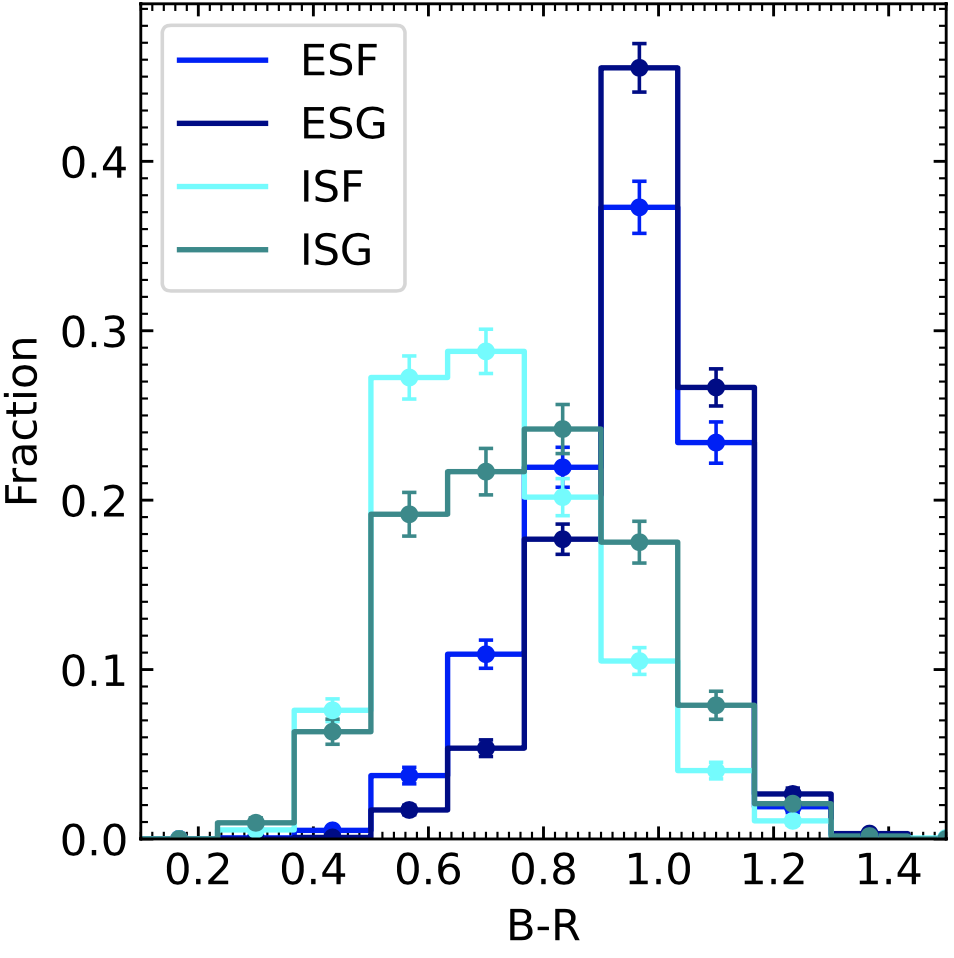}
  		}
  			\hspace{0.3cm}
  		\subfigure{
  			\includegraphics[width=0.3\linewidth]{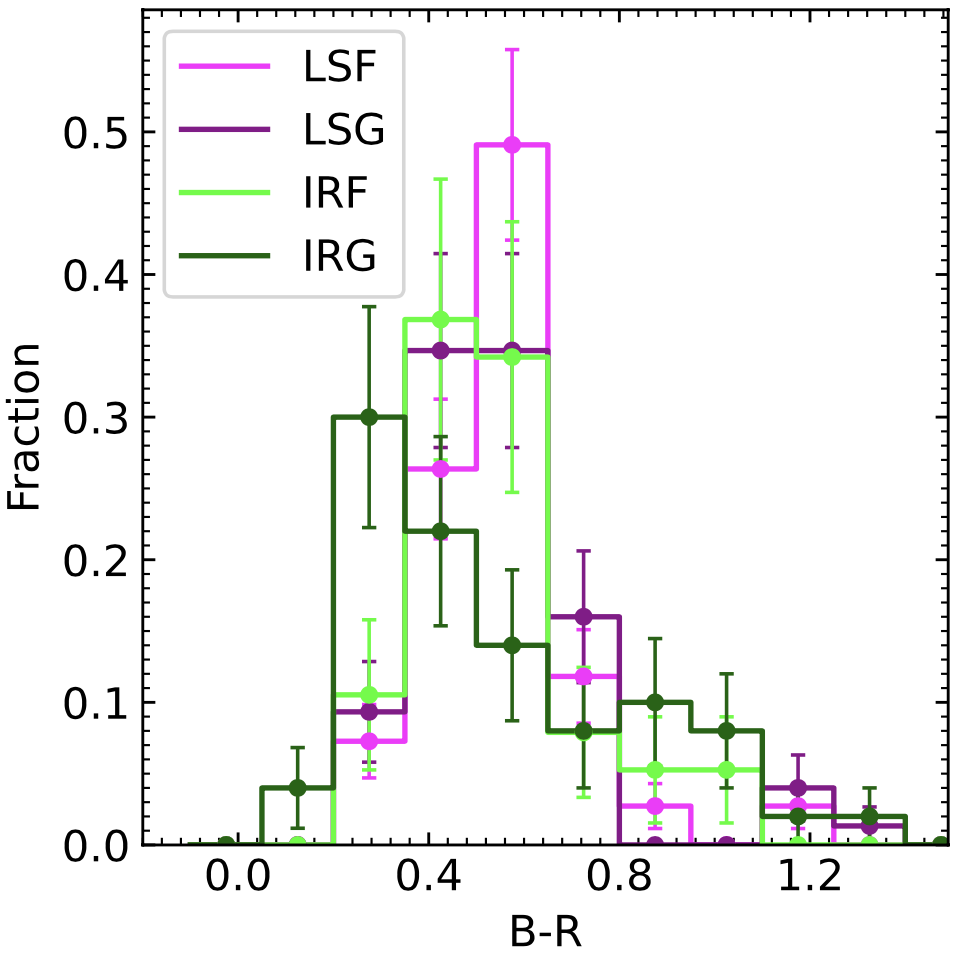}
  		}
  		\vspace{-0.2cm}
\caption{Comparison between field (light colour) and group (dark 
colour) galaxies' $B-R$  colour distributions. Left panel: Elliptical (red) 
and Lenticular (indigo); Middle panel: Early-type spiral (blue) and 
Intermediate-type spiral (cyan); Right panel: Late-type spiral (magenta) and 
Irregular (green) galaxies.}
\label{BR}
\end{figure}

\begin{figure}[h!]
  	\subfigure{
  		\includegraphics[width=0.3\linewidth]{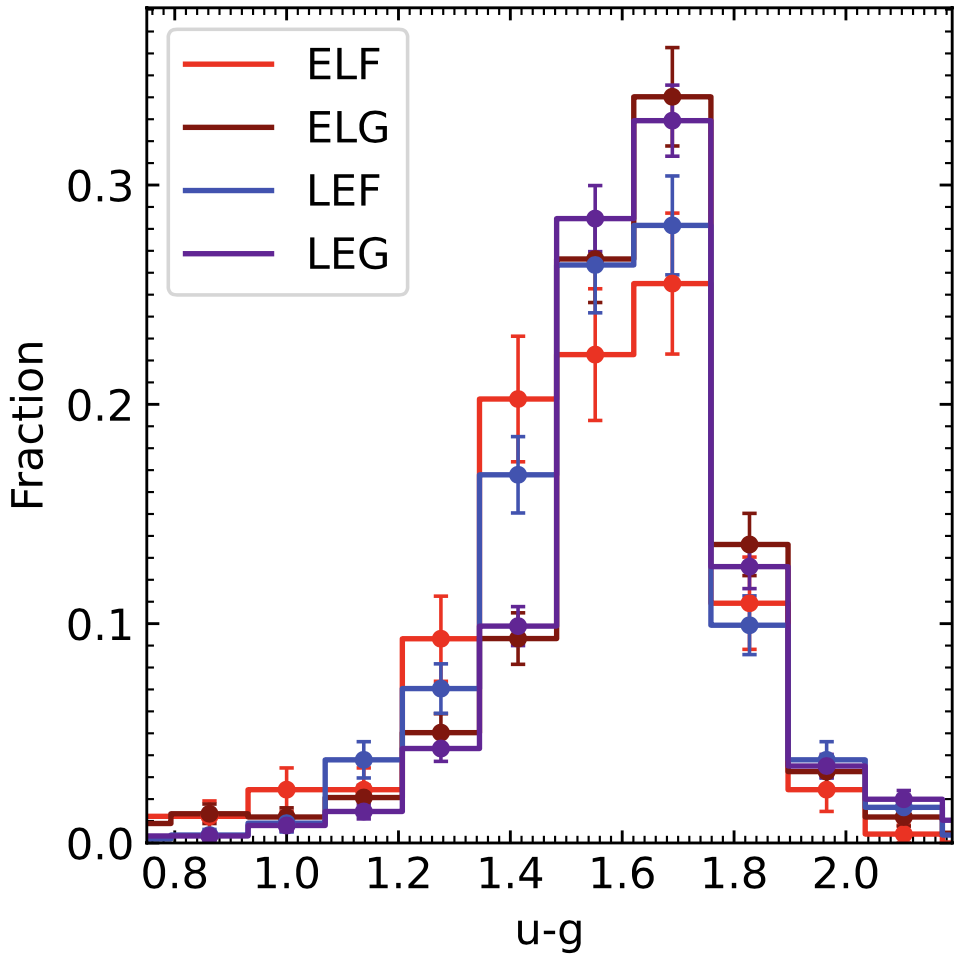}
  		}
  		\hspace{0.3cm}
  		\subfigure{
  			\includegraphics[width=0.3\linewidth]{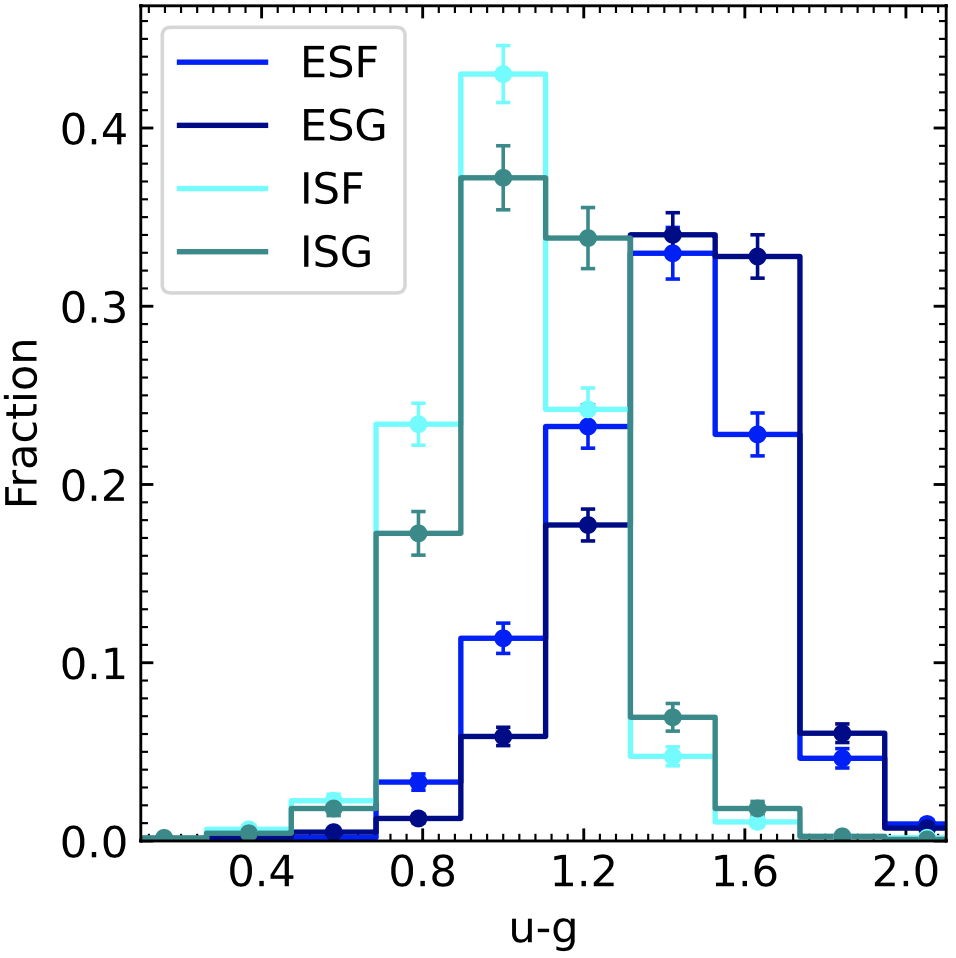}
  		}
  			\hspace{0.3cm}
  		\subfigure{
  			\includegraphics[width=0.3\linewidth]{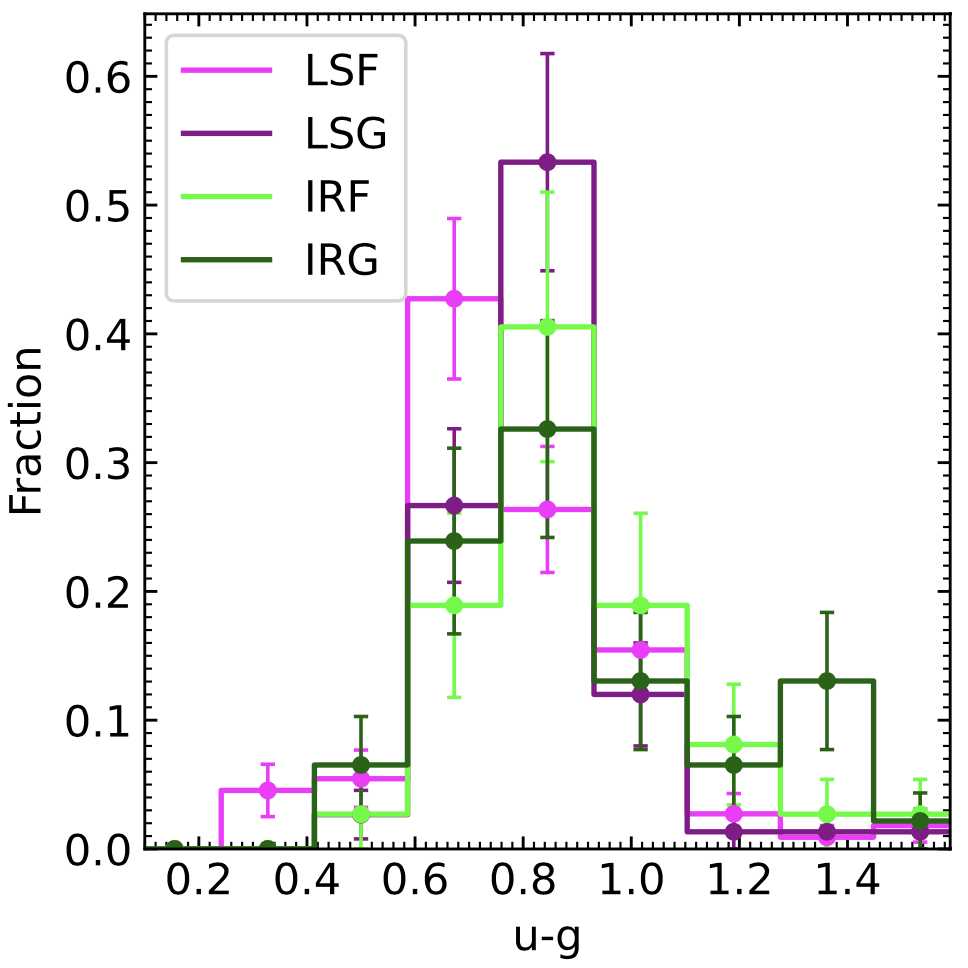}
  		}
        	\vspace{-0.2cm}
\caption{Comparison between field (light colour) and group (dark 
colour) galaxies' $u-g$ colour distributions. Left panel: Elliptical (red) 
and Lenticular (indigo); Middle panel: Early-type spiral (blue) and 
Intermediate-type spiral (cyan); Right panel: Late-type spiral (magenta) and 
Irregular (yellow) galaxies.}
\label{ug}
\end{figure}

\begin{figure}[h!]
  		\subfigure{
  			\includegraphics[width=0.3\linewidth]{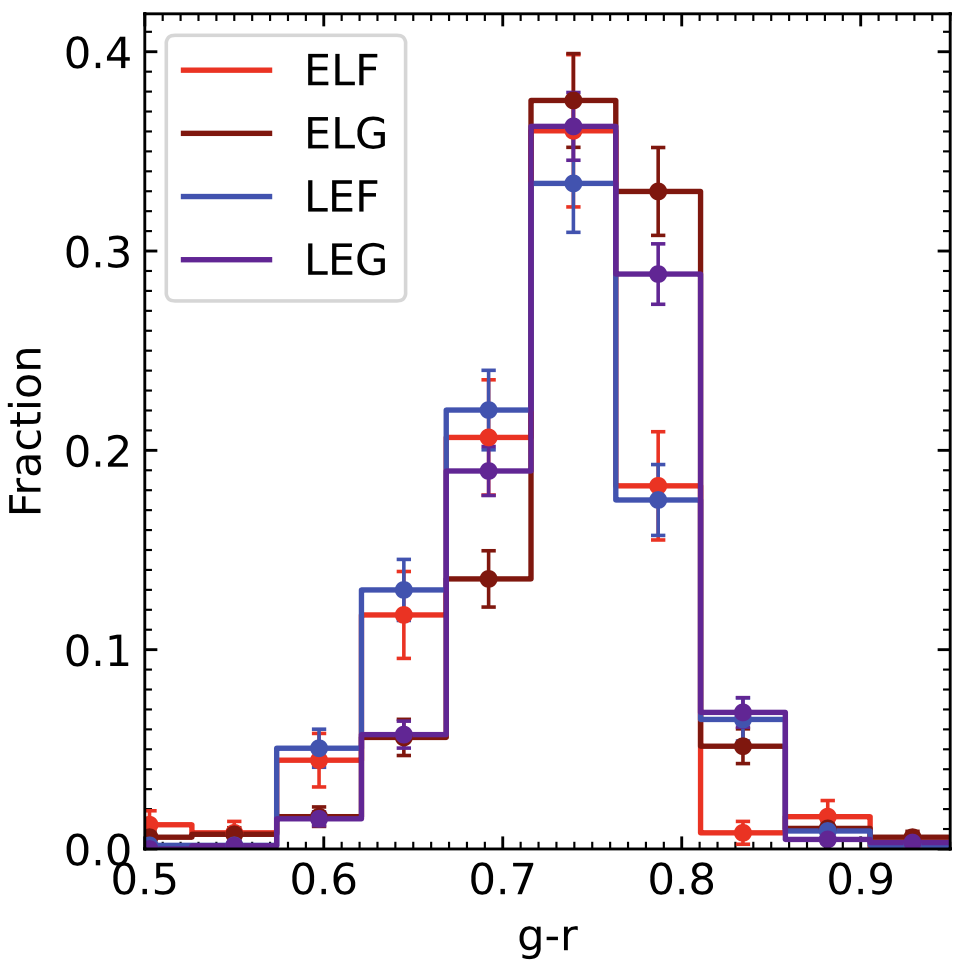}
  		}
  		\hspace{0.3cm}
  		\subfigure{
  			\includegraphics[width=0.3\linewidth]{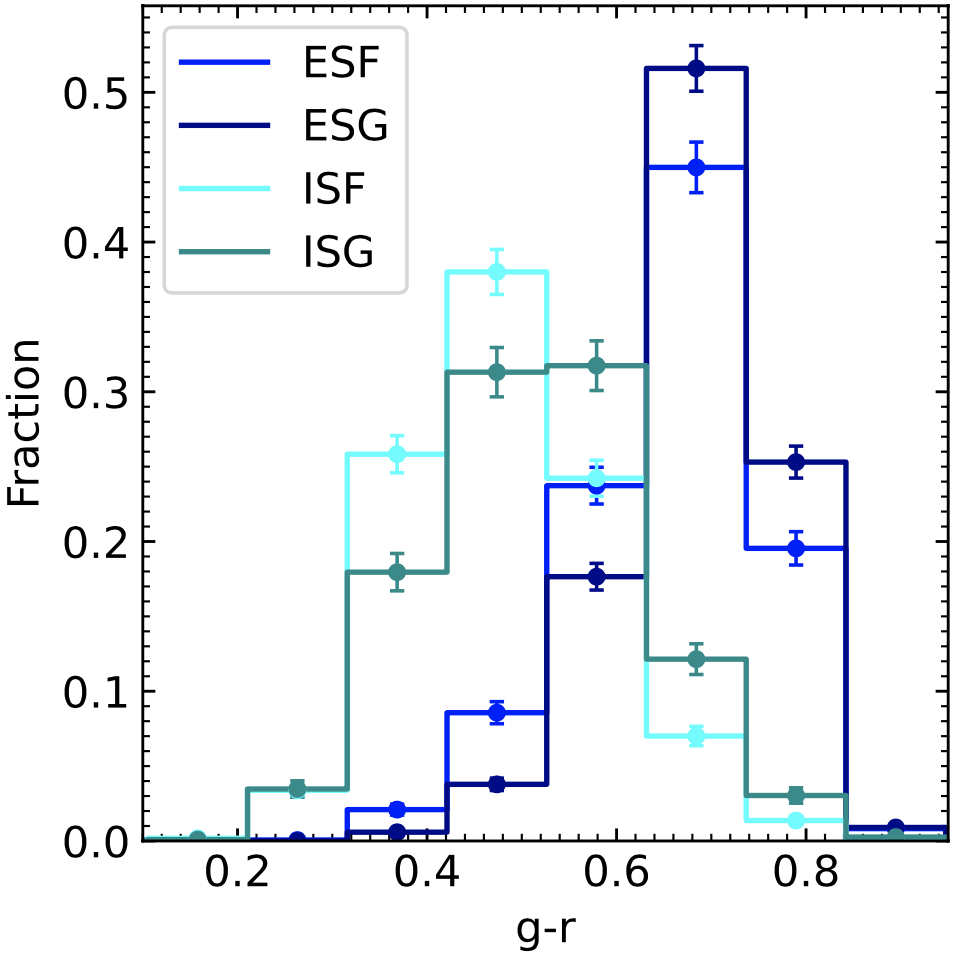}
  		}
  		\hspace{0.3cm}
  		\subfigure{
  			\includegraphics[width=0.3\linewidth]{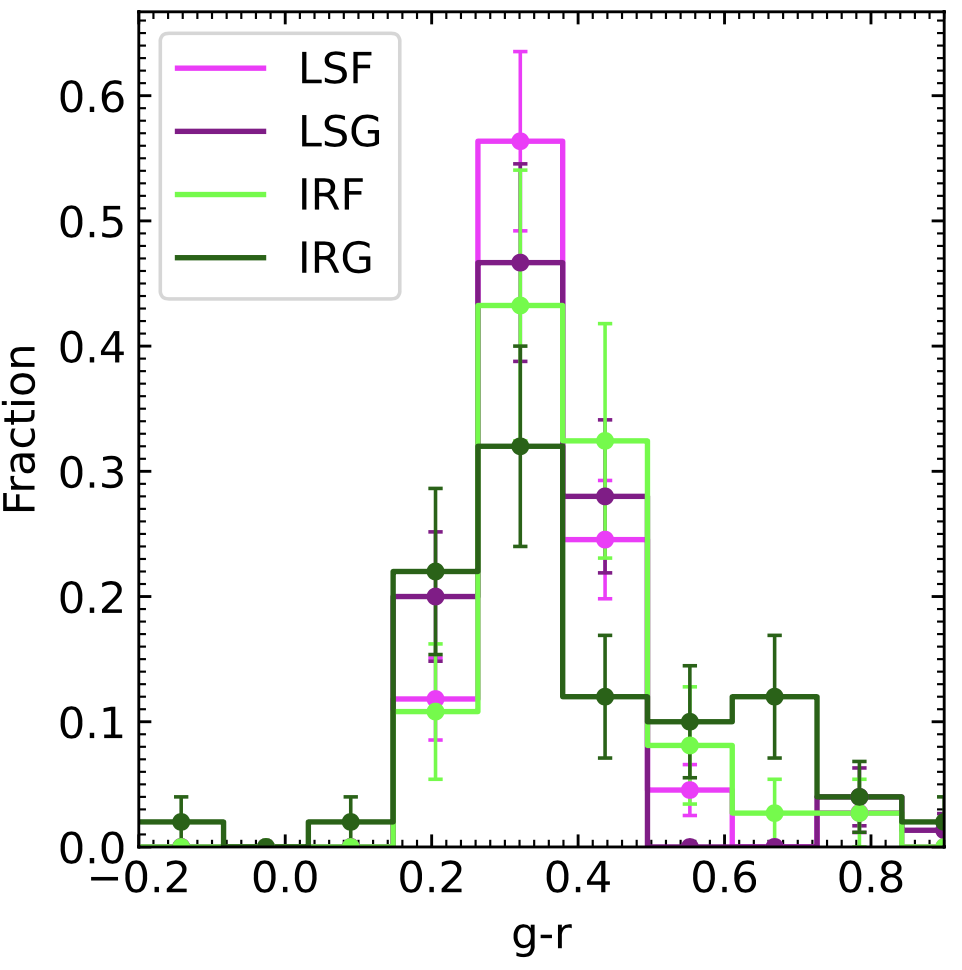}
  		}
  	        \vspace{-0.2cm}
\caption{Comparison between field (light colour) and group (dark 
colour) galaxies' $g-r$ colour distributions. Left panel: Elliptical (red) and 
Lenticular (green); Middle panel: Early-type spiral (blue) and 
Intermediate-type spiral (cyan); Right panel: Late-type spiral (magenta) and 
Irregular (green) galaxies.}
\label{gr}
\end{figure}

\begin{figure}[h!]
  		\subfigure{
  			\includegraphics[width=0.3\linewidth]{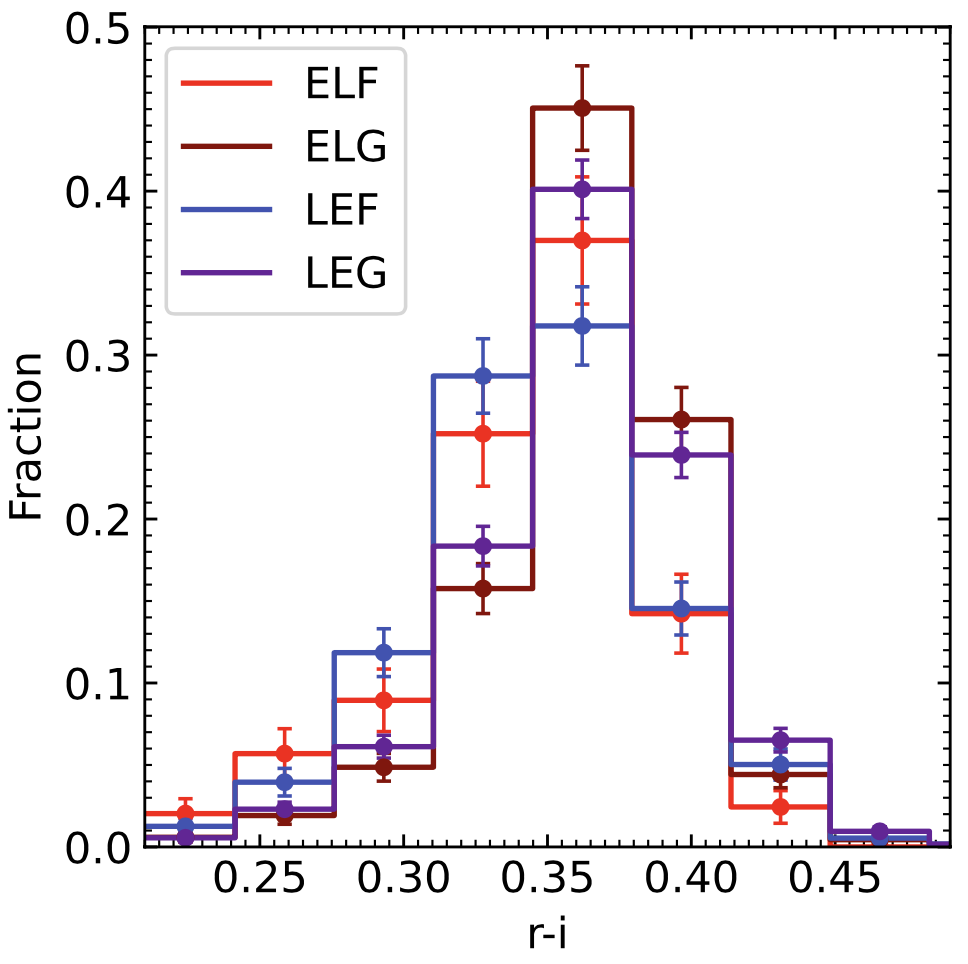}
  		}
  		\hspace{0.3cm}
  		\subfigure{
  			\includegraphics[width=0.3\linewidth]{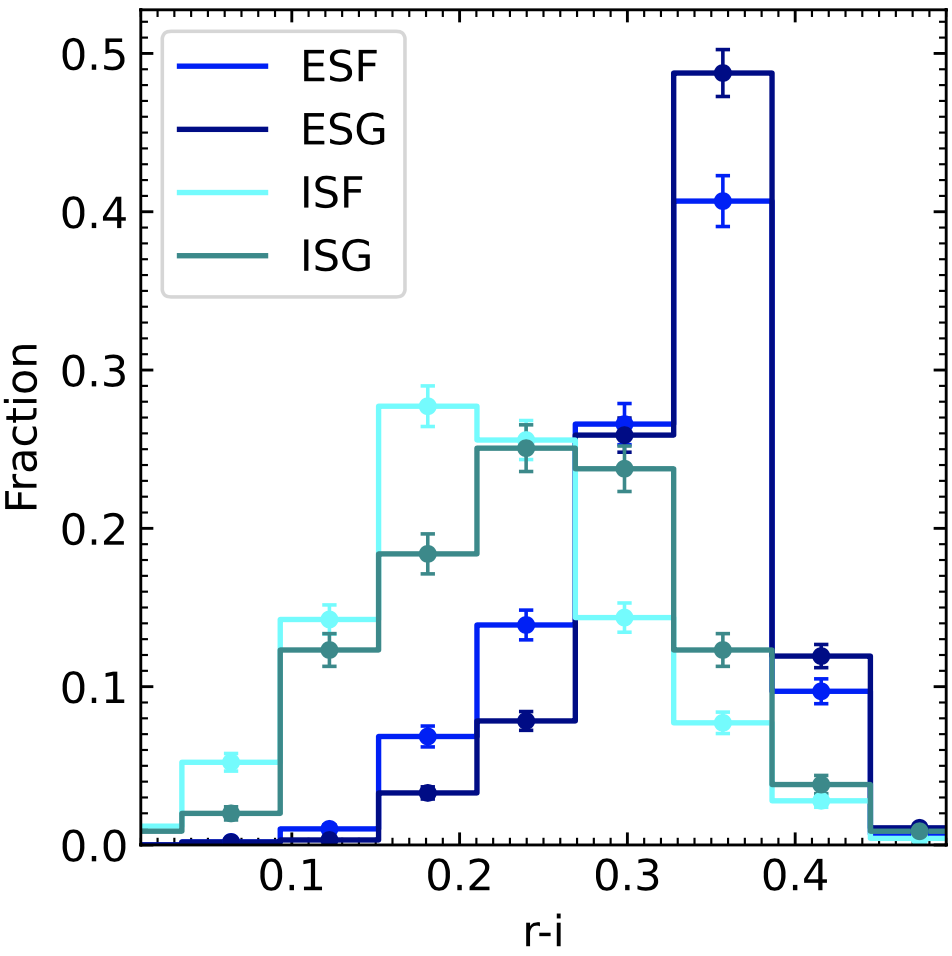}
  		}
  		\hspace{0.3cm}
  		\subfigure{
  			\includegraphics[width=0.3\linewidth]{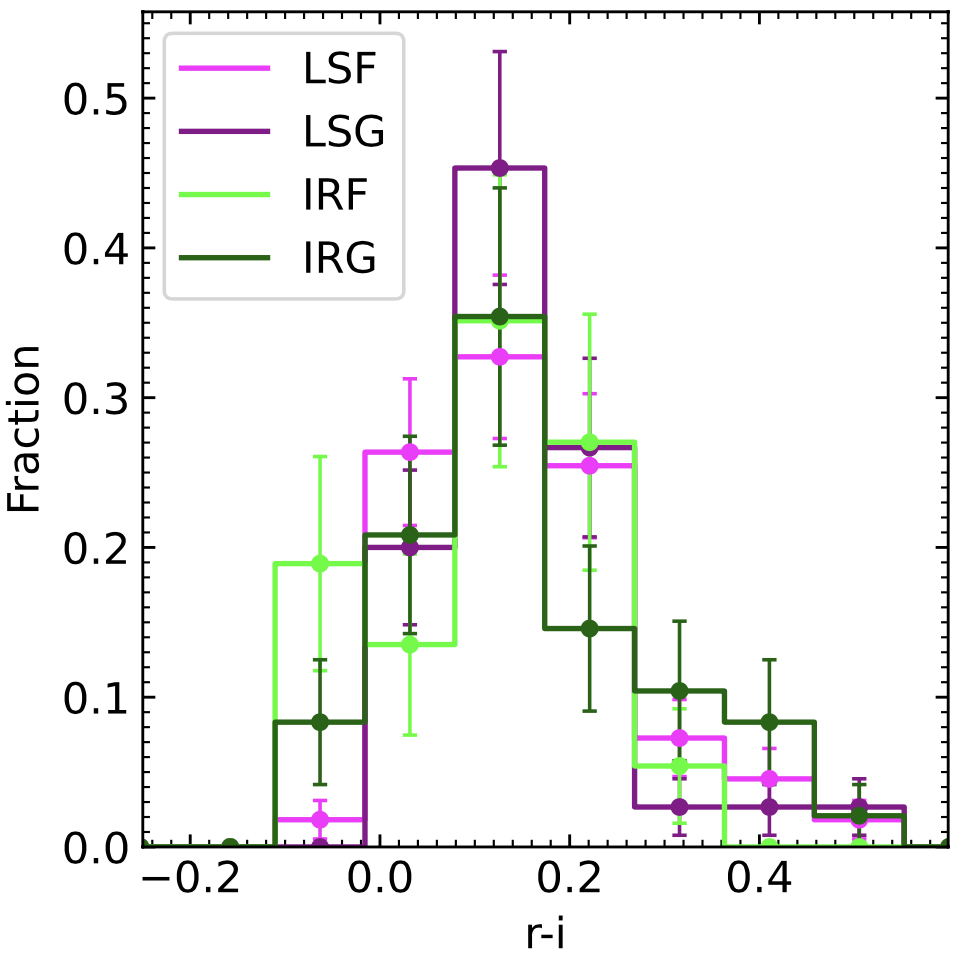}
  		}
  		\vspace{-0.2cm}
\caption{Comparison between field (light colour) and group (dark 
colour) galaxies' $r-i$ colour distributions. Left panel: Elliptical (red) 
and Lenticular (indigo); Middle panel: Early-type spiral (blue) and 
Intermediate-type spiral (cyan); Right panel: Late-type spiral (magenta) and 
Irregular (green) galaxies.}
\label{ri}
\end{figure}

\begin{figure}[h!]
  		\subfigure{
  			\includegraphics[width=0.3\linewidth]{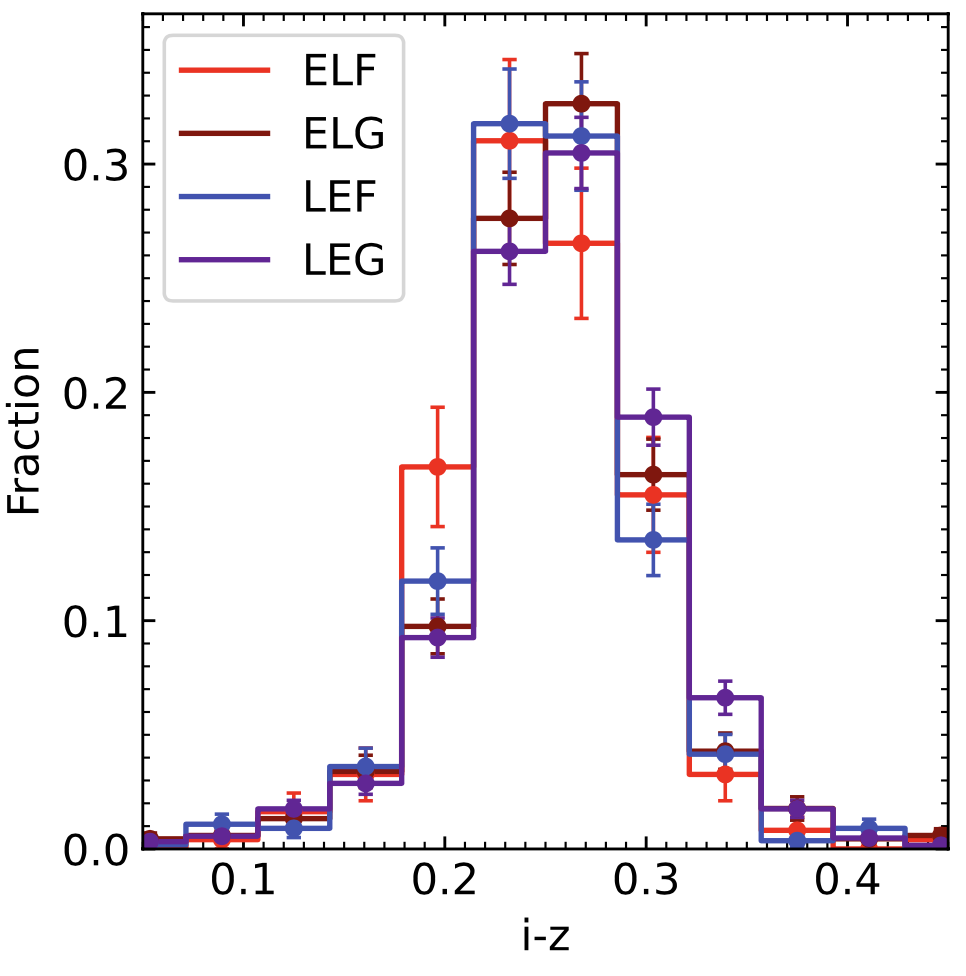}
  		}
  		\hspace{0.3cm}
  		\subfigure{
  			\includegraphics[width=0.3\linewidth]{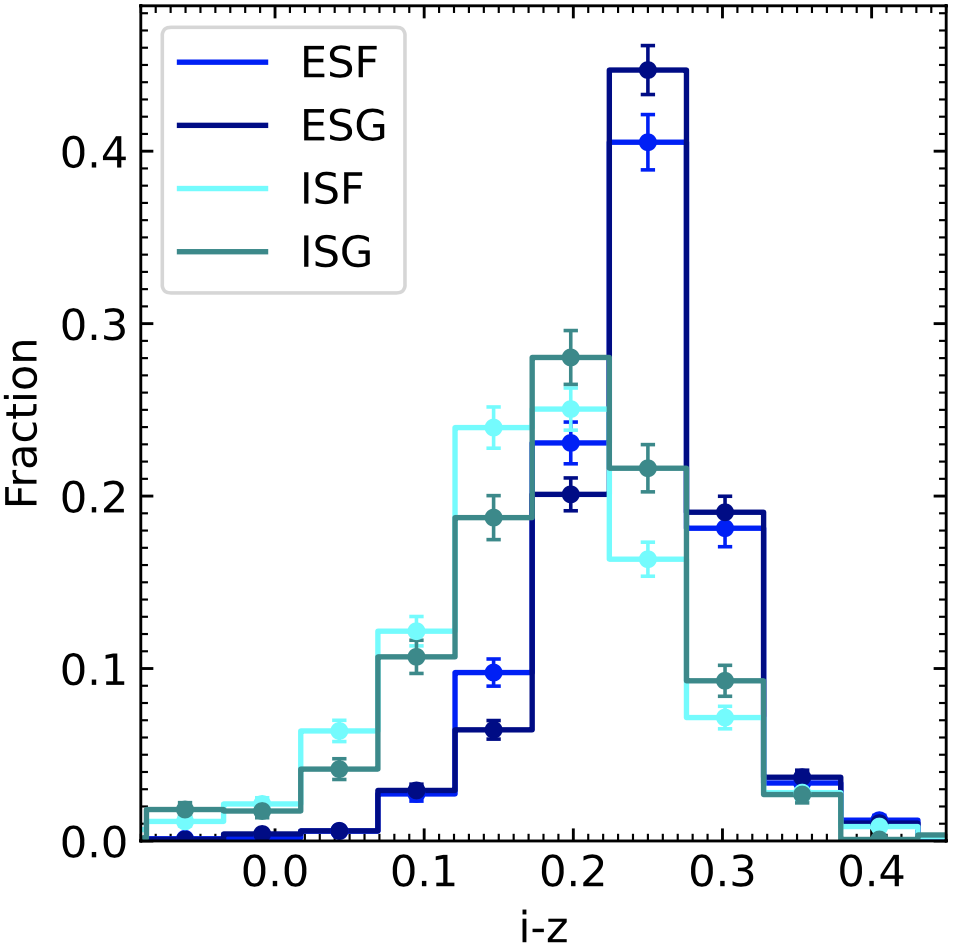}
  		}
  		\hspace{0.3cm}
  		\subfigure{
  			\includegraphics[width=0.3\linewidth]{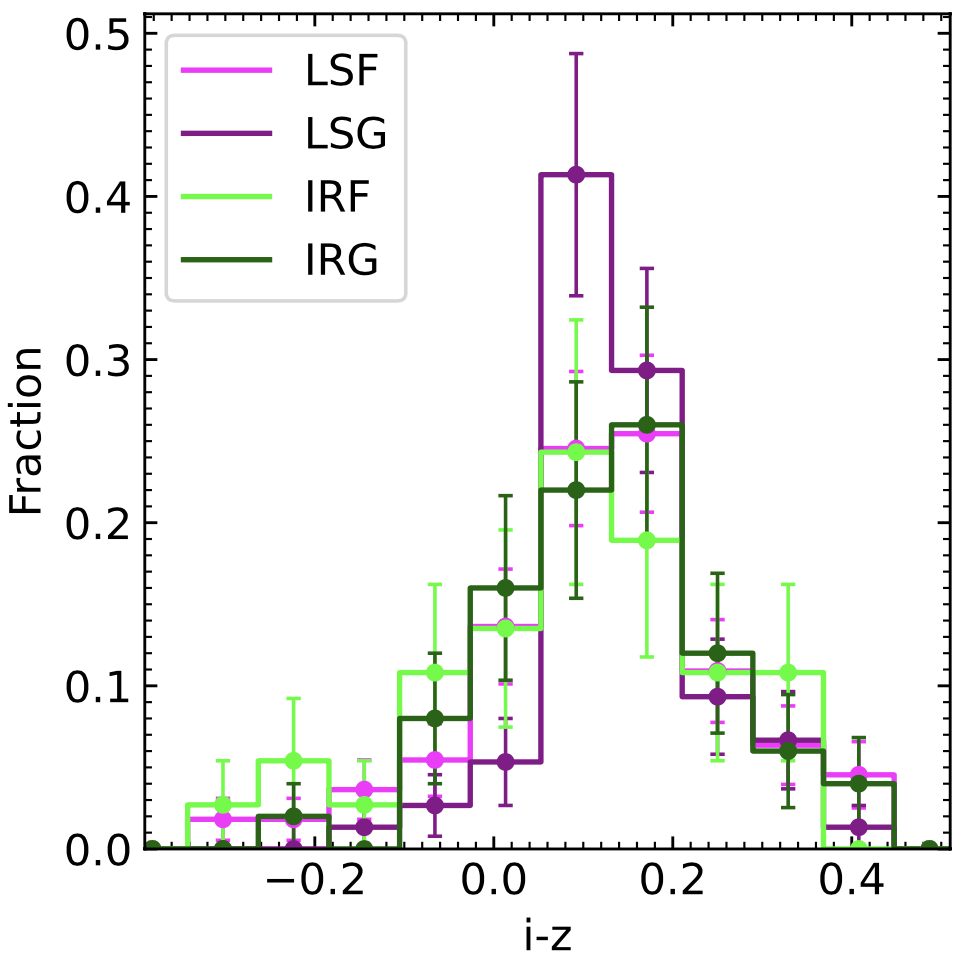}
  		}
  		\vspace{-0.2cm}
\caption{Comparison between field (light colour) and group (dark 
colour) galaxies' $i-z$ colour distributions. Left panel: Elliptical (red) 
and Lenticular (indigo); Middle panel: Early-type spiral (blue) and 
Intermediate-type spiral (cyan); Right panel: Late-type spiral (magenta) and 
Irregular (green) galaxies.}
\label{iz}
\end{figure}

\begin{figure}[h!]
	\subfigure{
		\includegraphics[width=0.31\linewidth]{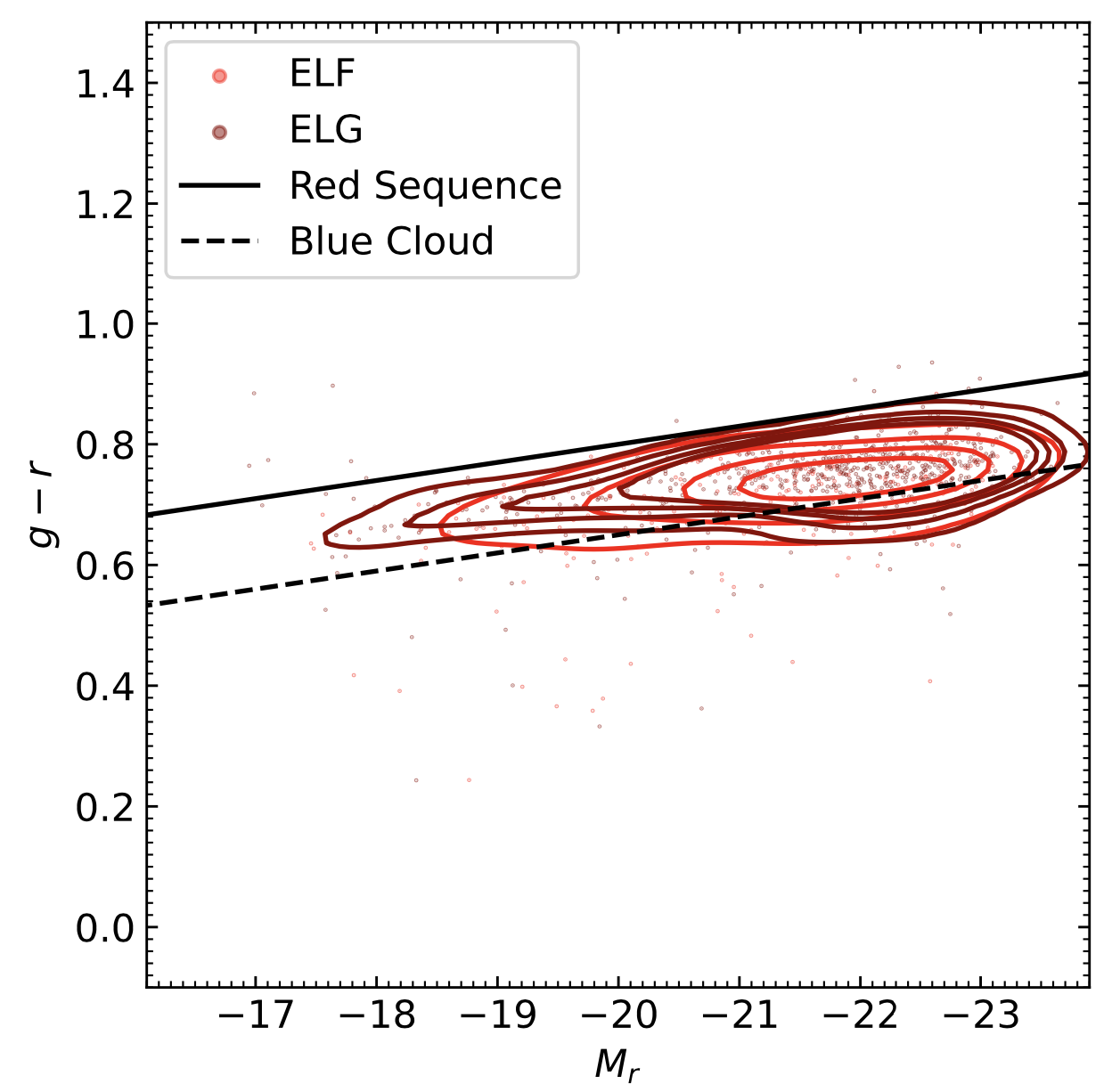}
	}
	\hspace{0.1cm}
	\subfigure{
		\includegraphics[width=0.31\linewidth]{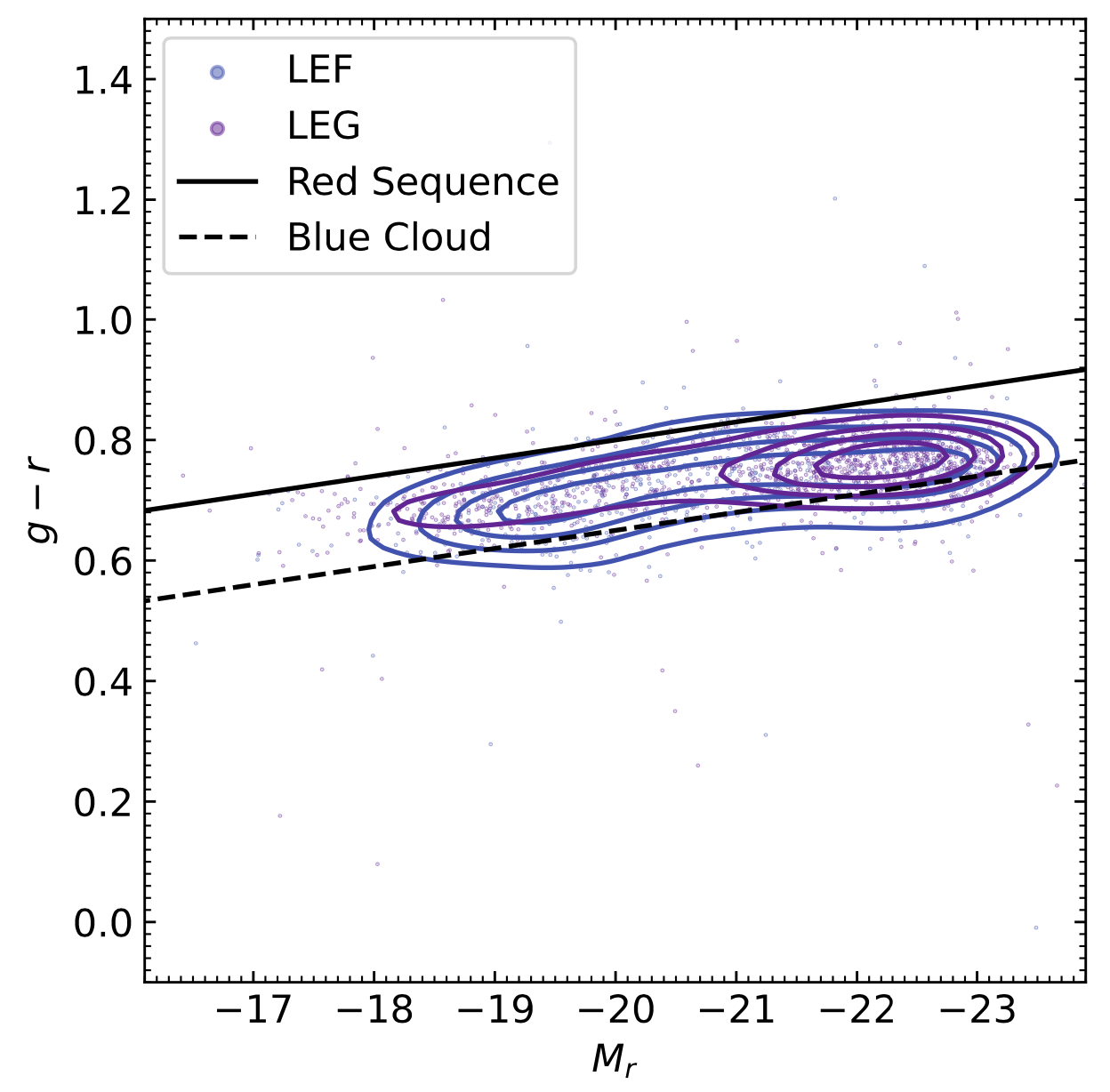}
	}
	\hspace{0.1cm}
	\subfigure{
		\includegraphics[width=0.31\linewidth]{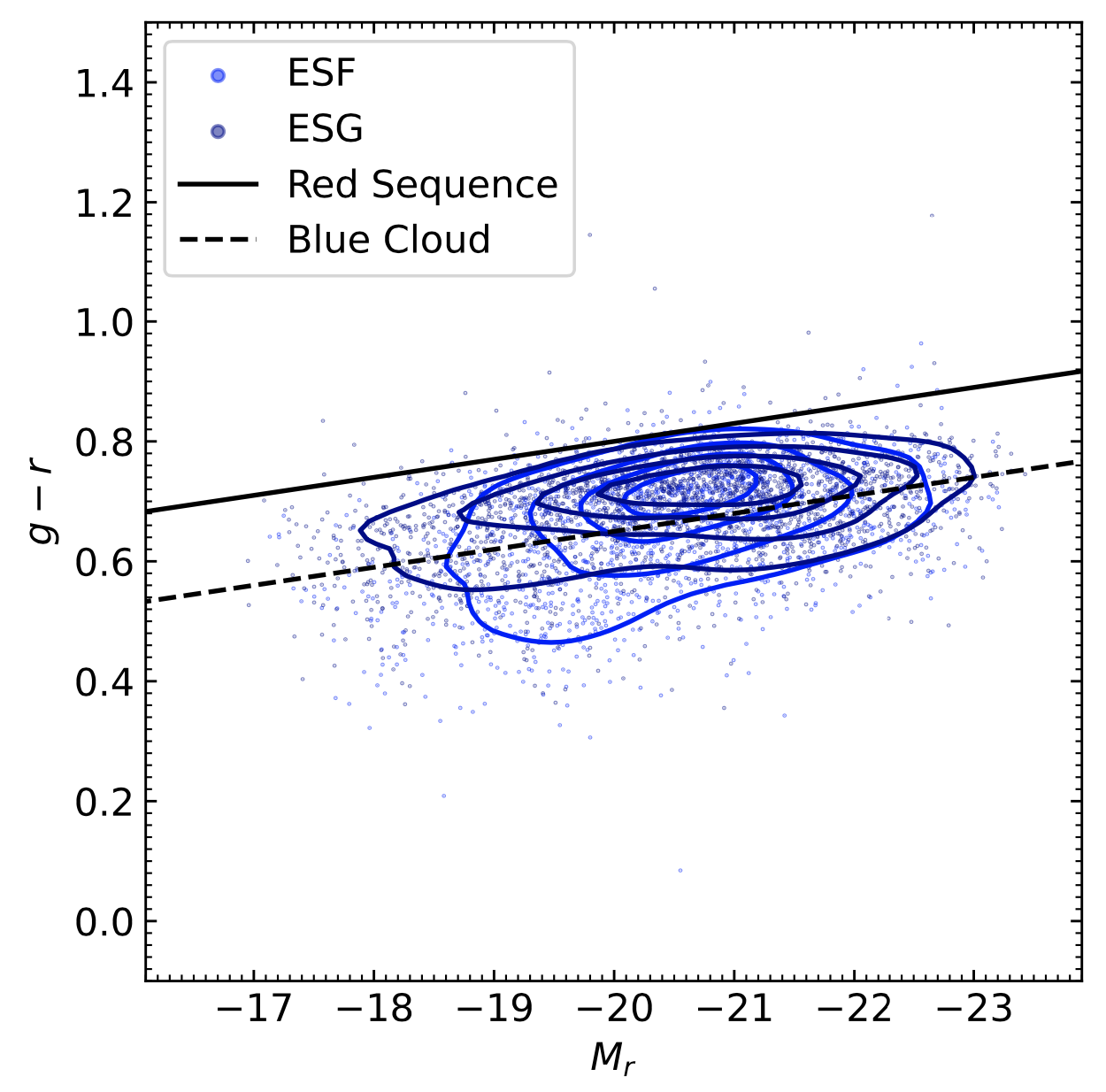}
	}
	\hspace{0.1cm}
	\subfigure{
		\includegraphics[width=0.31\linewidth]{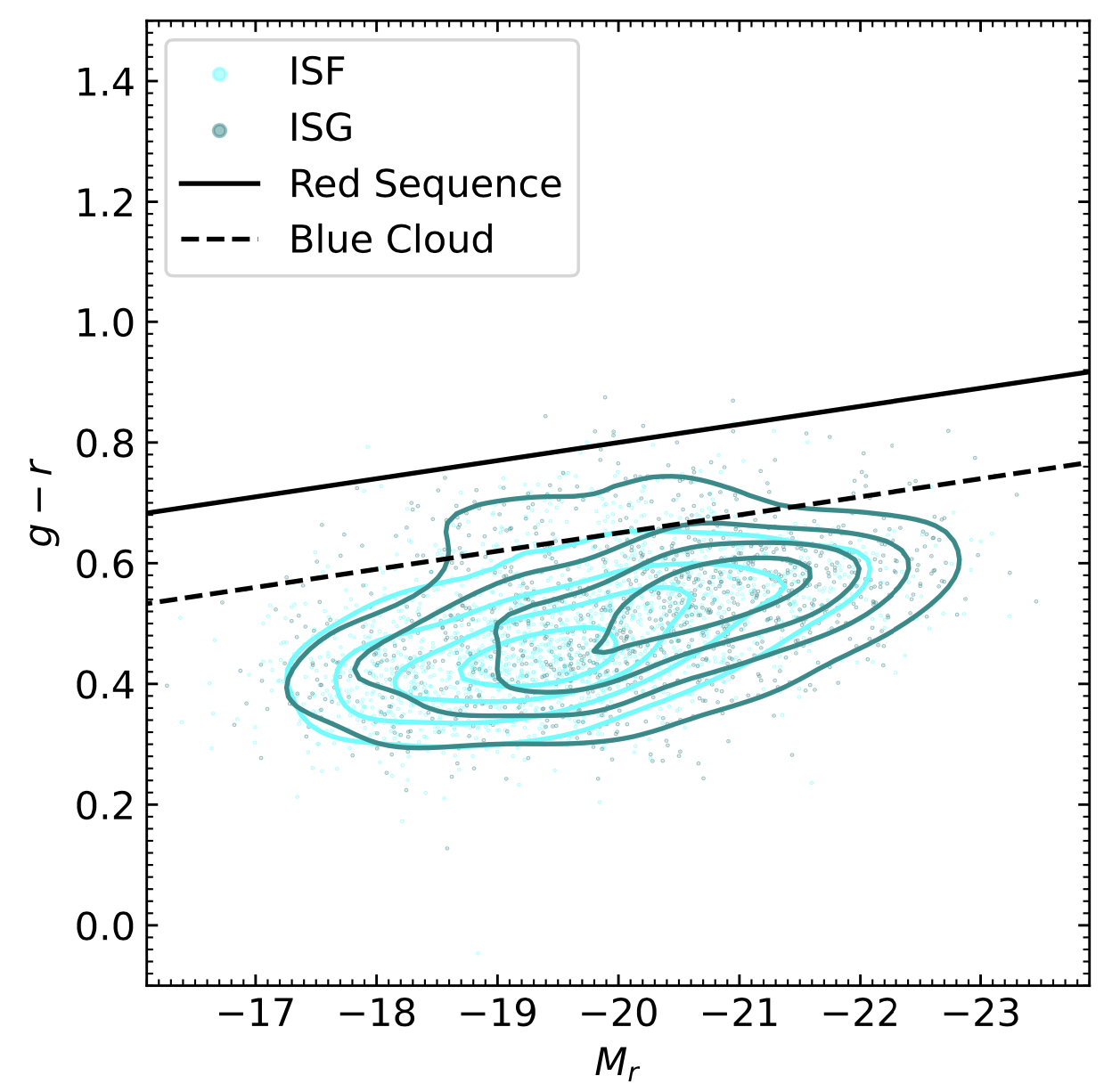}
	}
	\hspace{0.1cm}
	\subfigure{
		\includegraphics[width=0.31\linewidth]{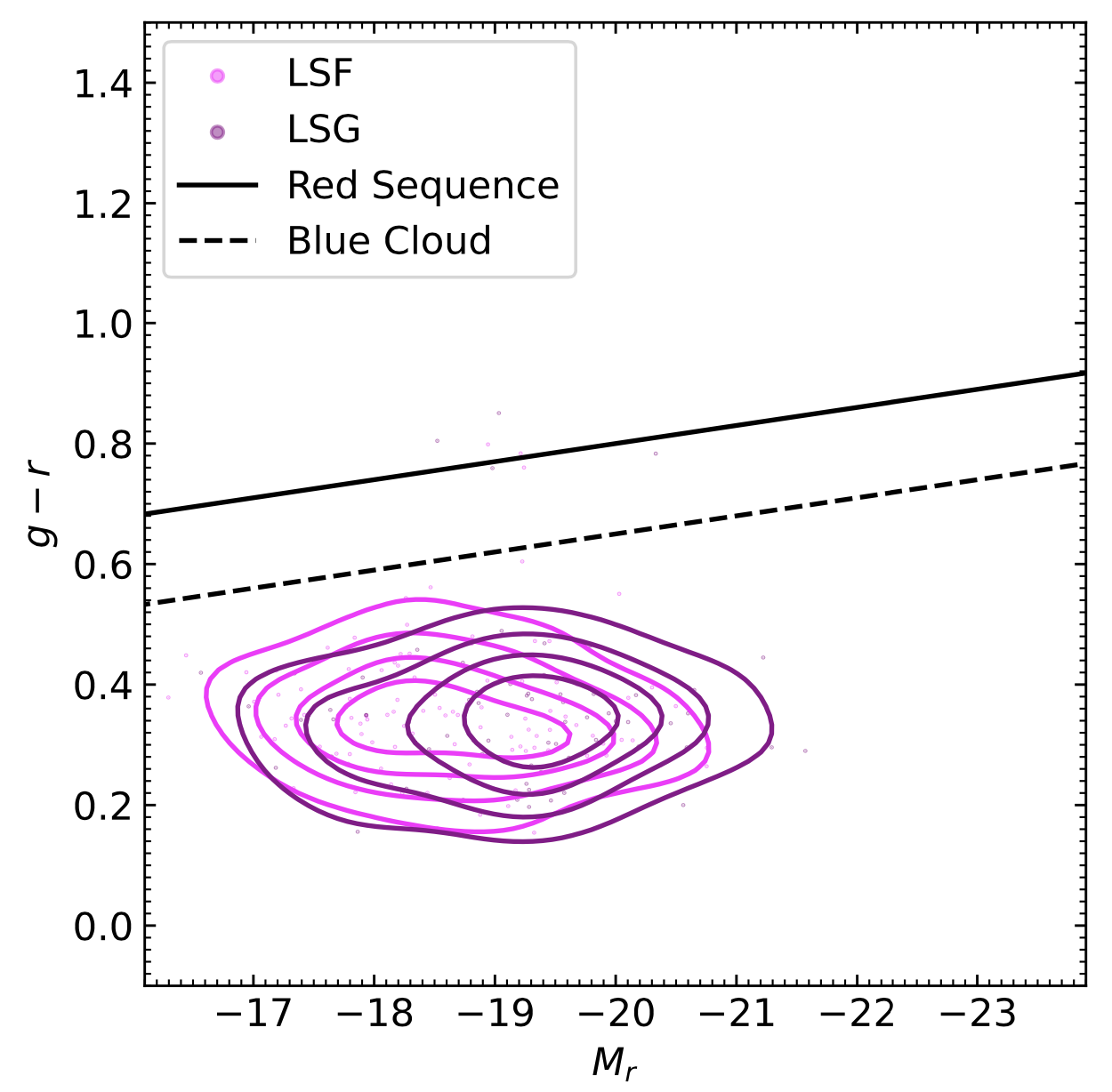}
	}
	\hspace{0.1cm}
	\subfigure{
		\includegraphics[width=0.31\linewidth]{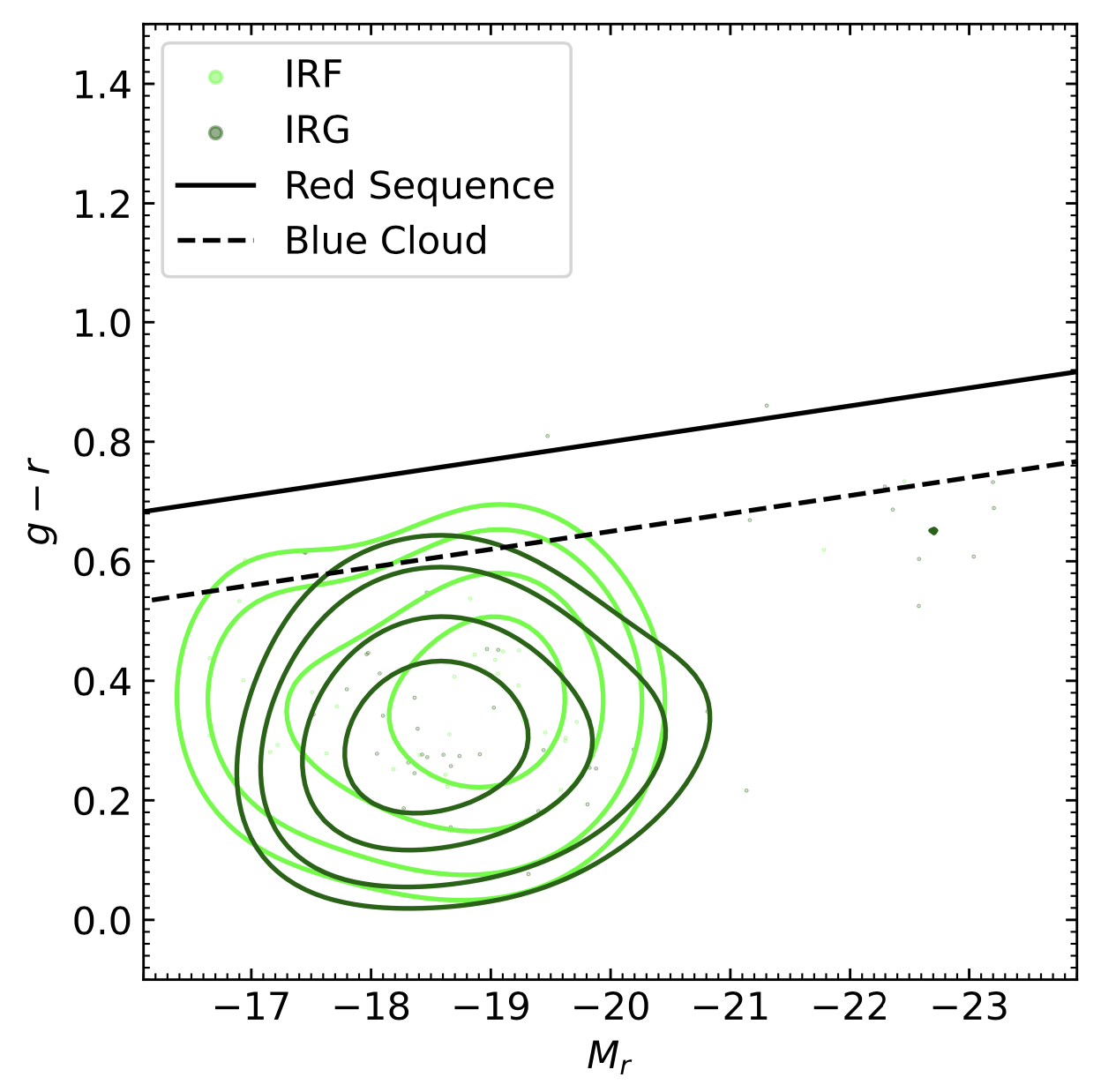}
	}
	\hspace{-0.3cm}
\caption{Comparison of colour-magnitude planes between field (light 
colour) and group (dark colour) types for EL, LE, ES, IS, LS and IR galaxies. 
The contours show the 80\%, 60\%, 40\% and 20\% of probability distribution.}
	\label{CM}
\end{figure}

In this section, we compare the distributions of galaxy colours 
($B-V$, $B-R$, $u-g$, $g-r$, $r-i$, $i-z$) between the field and group 
galaxies as shown by Figs.~\ref{BV}, \ref{BR}, \ref{ug}, \ref{gr}, \ref{ri} 
and \ref{iz} respectively. The Johnson $B$, $V$ and $R$ filters were computed 
by pypipe3D from the MaNGA data cubes \cite{sanchez2022sdss}, following the 
Vega photometric system and employing the filter parameters specified in 
Refs.\ \cite{fukugita1995galaxy,blanton2007k} the Vega system was 
converted to AB system, while the $u-g$, $g-r$, $r-i$ and $i-z$ were 
obtained from the NASA-Sloan Atlas (NSA) \cite{blanton2011improved}. 
The use of both MANGA IFU and SDSS photometry is very important 
since it enables us to assess if the observed difference in colour between 
field and group galaxies is similar for both photometry, keeping in mind that 
there are aperture differences and hence we avoid mixing them during the 
analysis. We performed the Kolmogorov-Smirnov (KS) test 
\cite{hodges1958significance,
harari2009kolmogorov} as shown in Table \ref{KS} and Anderson–Darling (AD) 
\cite{anderson1952asymptotic,pettitt1976two,scholz1987k,
babu2006astrostatistics} statistical test as shown in Table \ref{AD}, to 
assess the degree of similarity of field and group galaxies' colours 
distributions. Further, we compare the colour-magnitude diagrams 
between field (light) and group(dark) galaxies for EL (red), LE (green), 
ES (blue), IS (cyan), LS (magenta) and IR (yellow) as shown by 
Fig.~\ref{CM} wherein the Blue cloud and Red sequence galaxies are defined by 
criteria \eqref{CM1} and \eqref{CM2} respectively, obtained from 
Refs.\ \cite{blanton2005properties,dhiwar2023witnessing} as follows:
\begin{align}
(g-r)<0.65-0.03 \times\left(M_r+20\right),
\label{CM1}\\[5pt]
(g-r)>0.80-0.03 \times\left(M_r+20\right),
  		\label{CM2}
\end{align}
where $M_r$ is the $r$-band absolute magnitude.  		
  	

\begin{table}[!h]
\centering
\caption{The KS statistics with their corresponding p-values (in brackets) 
for the colours of field and group types of Elliptical (EL), Lenticular (LE), 
Early-type (ES), Intermediate-type (IS), Late-type (LS) spirals, and Irregular 
(IR) galaxies.}
  		\vspace{8pt}
  		\setlength{\tabcolsep}{0.3pc}
  		\scalebox{1.0}{
  			\begin{tabular}{ccccccc}
  				\toprule
  				\toprule
  				& \multicolumn{6}{c}{KS statistics (p-values)} \\
  				\cmidrule(lr){2-7}
  				Colour & EL & LE & ES & IS & LS & IR \\
  				{(1)} & {(2)} & {(3)} & {(4)} & {(5)} & {(6)} & {(7)} \\
  				\midrule
  				$B-V$ & \num{0.20} (\num{8.52e-07}) & \num{0.14} (\num{6.89e-07}) & \num{0.14} (\num{8.97e-16}) & 
  				\num{0.15} (\num{2.20e-14}) & \num{0.18} (\num{0.09}) & \num{0.23} (\num{0.17}) \\[2pt]
  				$B-R$ & \num{0.21} (\num{1.24e-07}) & \num{0.16} (\num{7.45e-09}) & \num{0.13} (\num{1.39e-13}) & 
  				\num{0.17} (\num{6.73e-17}) & \num{0.14} (\num{0.27})&\num{0.23} (\num{0.15}) \\[2pt]
  				$u-g$ & \num{0.20} (\num{7.65e-07}) & \num{0.17} (\num{4.02e-10}) & \num{0.15} (\num{1.65e-19}) &
  				\num{0.13} (\num{1.33e-10}) & \num{0.24} (\num{0.08})& \num{0.16} (\num{0.60})\\[2pt]
  				$g-r$ & \num{0.23} (\num{1.03e-08}) & \num{0.16} (\num{7.65e-09}) & \num{0.14} (\num{1.51e-15}) &
  				\num{0.16} (\num{1.17e-15}) &\num{0.08} (\num{0.90}) &\num{0.29} (\num{0.05}) \\[2pt]
  				
  				$r-i$ & \num{0.24} (\num{6.48e-10}) & \num{0.20} (\num{7.00e-14}) & \num{0.12} (\num{7.06e-13}) &
  				\num{0.17} (\num{9.58e-18}) & \num{0.12} (\num{0.49})&\num{0.17} (\num{0.52}) \\[2pt]
  				
  				$i-z$ & \num{0.11} (\num{1.56e-02}) & \num{0.11} (\num{3.44e-04}) & \num{0.06} (\num{8.61e-04}) &
  				\num{0.11} (\num{1.35e-07}) & \num{0.20} (\num{0.05})&\num{0.15} (\num{0.67}) \\
  				\bottomrule
  		\end{tabular}}
  		\label{KS}
\end{table}
  	
\begin{table}[h!]
  	\centering
\caption{The AD statistics with their corresponding p-values (in brackets) 
for the colours of field and group types of Elliptical (EL), Lenticular (LE), 
Early-type (ES), Intermediate-type (IS), Late-type spirals (LS), and Irregular 
(IR) galaxies.}
  		\vspace{8pt}
  		\setlength{\tabcolsep}{0.3pc}
  		\scalebox{1.0}{
  			\begin{tabular}{ccccccc}
  				\toprule
  				\toprule
  				& \multicolumn{6}{c}{AD statistics (p-values)} \\
  				\cmidrule(lr){2-7}
  				Colour & EL & LE & ES & IS & LS & IR \\
  				{(1)} & {(2)} & {(3)} & {(4)} & {(5)} & {(6)} & {(7)} \\
  				\midrule
  				$B-V$ & \num{24.68} (\num{1.00 e-05}) & \num{23.76} (\num{1.00e-05}) & \num{60.15} (\num{1.00e-05}) & 
  				\num{44.17} (\num{1.00e-05}) & \num{1.01} (\num{0.13}) & \num{1.73} (\num{0.06}) \\[2pt]
  				$B-R$ & \num{25.71} (\num{1.00e-05}) & \num{26.19} (\num{1.00e-05}) & \num{43.72} (\num{1.00e-05}) & 
  				\num{48.61} (\num{1.00e-05}) & \num{0.17} (\num{0.25}) &\num{1.55} (\num{0.07}) \\[2pt]
  				$u-g$ & \num{13.71} (\num{1.00e-05}) & \num{19.15} (\num{1.00e-05}) & \num{59.93} (\num{1.00e-05}) & 
  				\num{29.81} (\num{1.00e-05}) & \num{3.09} (\num{0.02}) &\num{0.37} (\num{0.25}) \\[2pt]
  				$g-r$ & \num{32.94} (\num{1.00e-05}) & \num{32.79} (\num{1.00e-05}) & \num{49.62} (\num{1.00e-05}) & 
  				\num{46.25} (\num{1.00e-05}) & \num{0.75} (\num{0.25}) &\num{2.03} (\num{0.05}) \\[2pt]
  				
  				$r-i$ & \num{31.40} (\num{1.00e-05}) & \num{38.70} (\num{1.00e-05}) & \num{40.48} (\num{1.00e-05}) & 
  				\num{51.02} (\num{1.00e-05}) & \num{0.17} (\num{0.25}) &\num{0.01} (\num{0.25}) \\[2pt]
  				
  				$i-z$ & \num{2.71} (\num{2.53e-04}) & \num{9.65} (\num{1.00e-05}) & \num{7.02} (\num{1.00e-05}) & 
  				\num{14.19} (\num{1.00e-05}) & \num{1.59} (\num{0.07}) &\num{0.11} (\num{0.25}) \\
  				\bottomrule
\end{tabular}}
\label{AD}
\end{table}

\begin{figure}[h!]
  		\subfigure{
  				\includegraphics[width=0.3\linewidth]{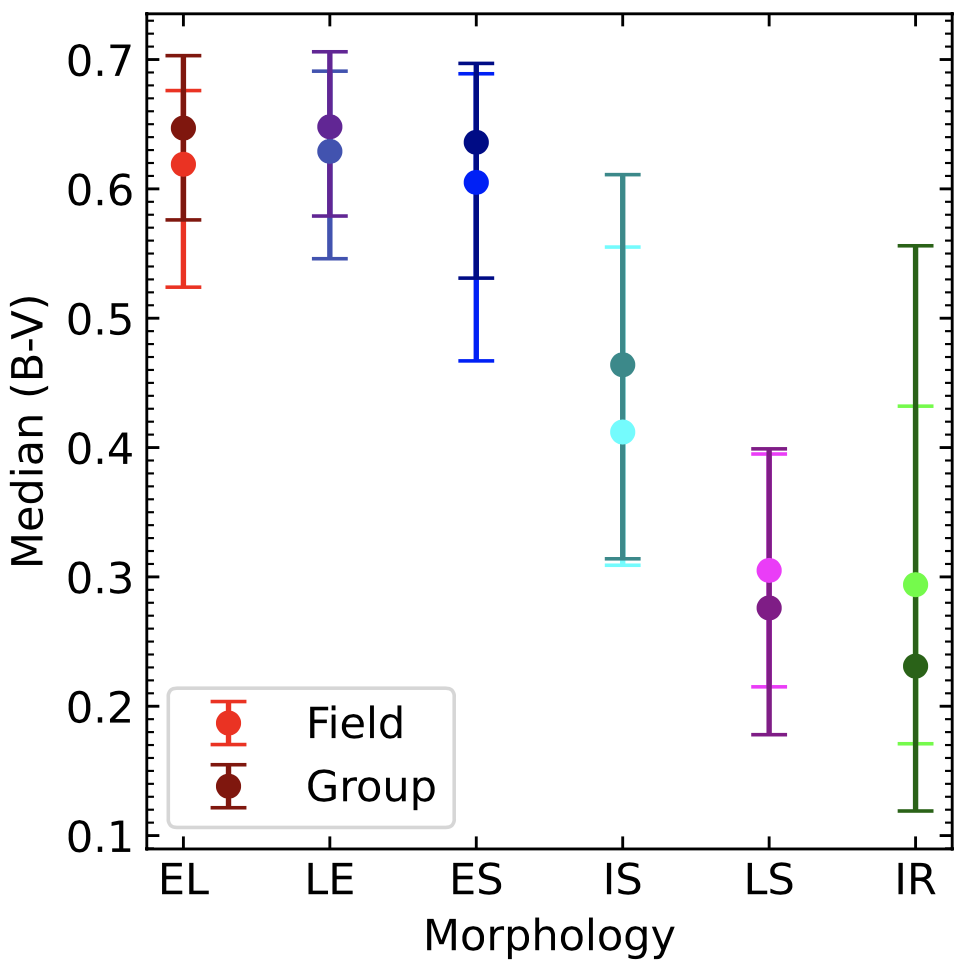}
  			}
  			\hspace{0.3cm}
  		\subfigure{
  				\includegraphics[width=0.3\linewidth]{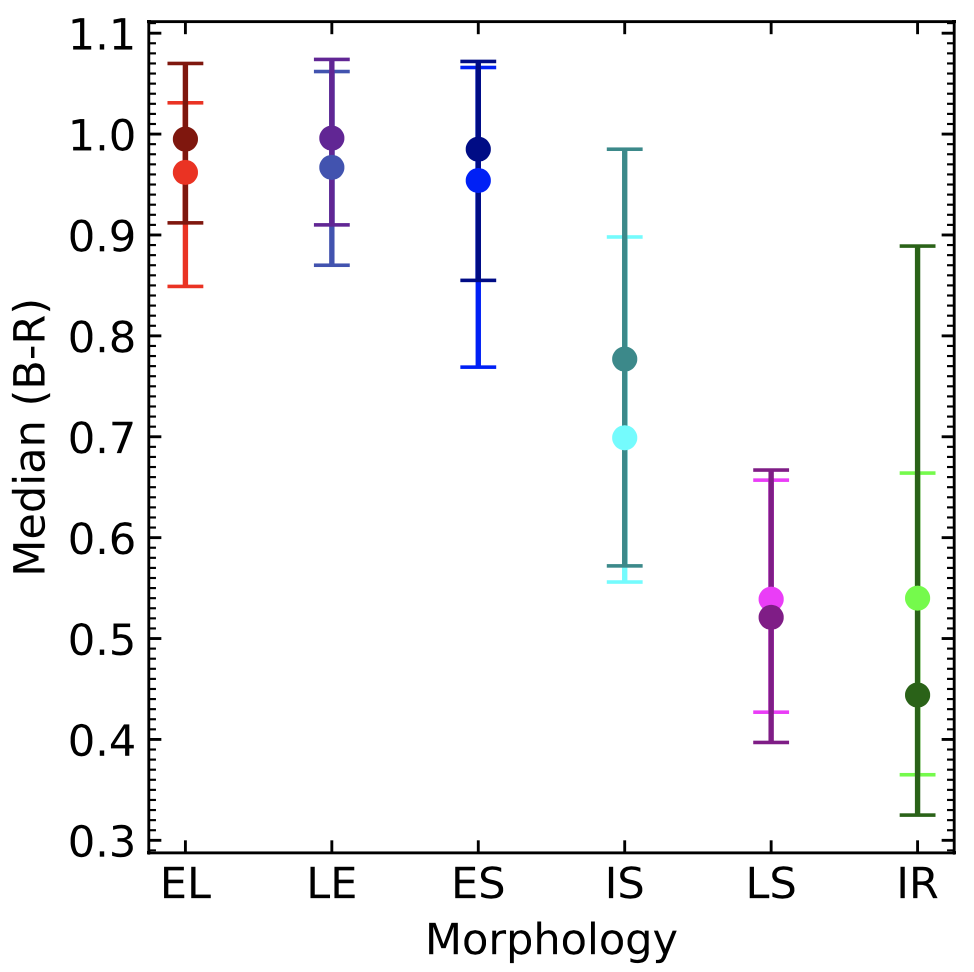}
  			}
  			\hspace{0.3cm}
  		\subfigure{
  			\includegraphics[width=0.3\linewidth]{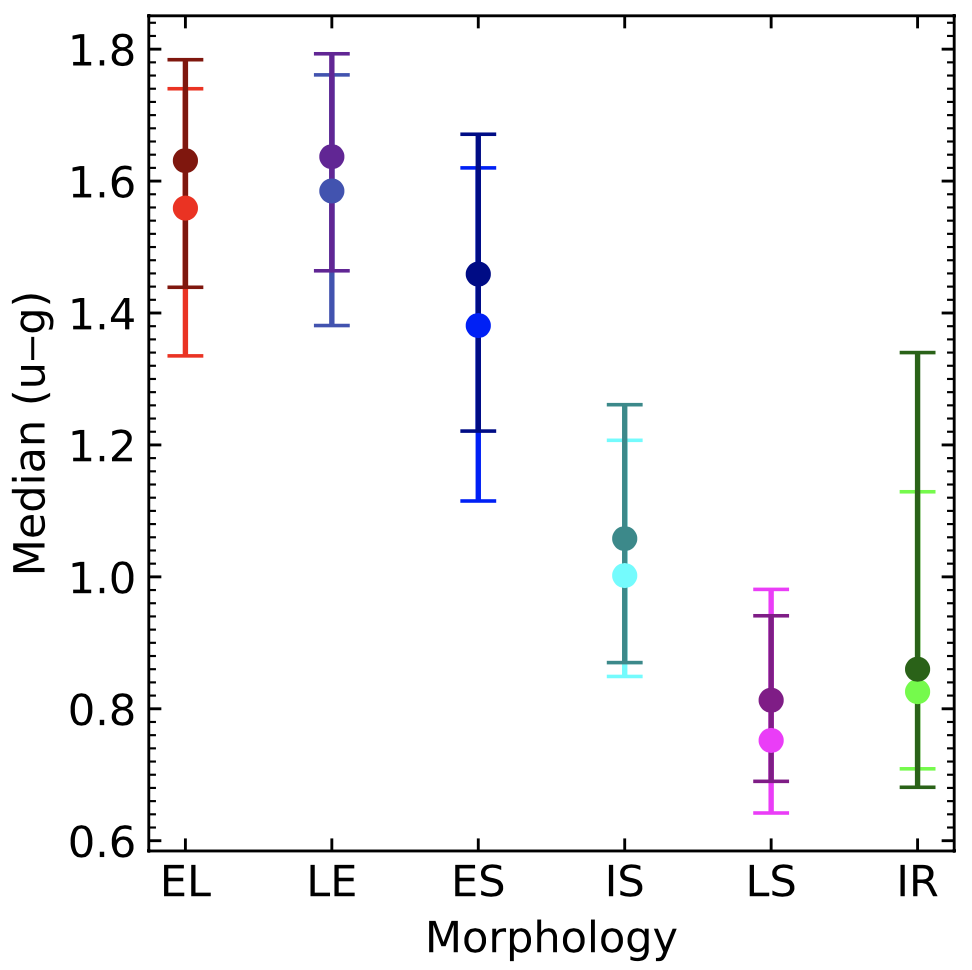}
  		}
  		\hspace{0.3cm}
  		\subfigure{
  			\includegraphics[width=0.3\linewidth]{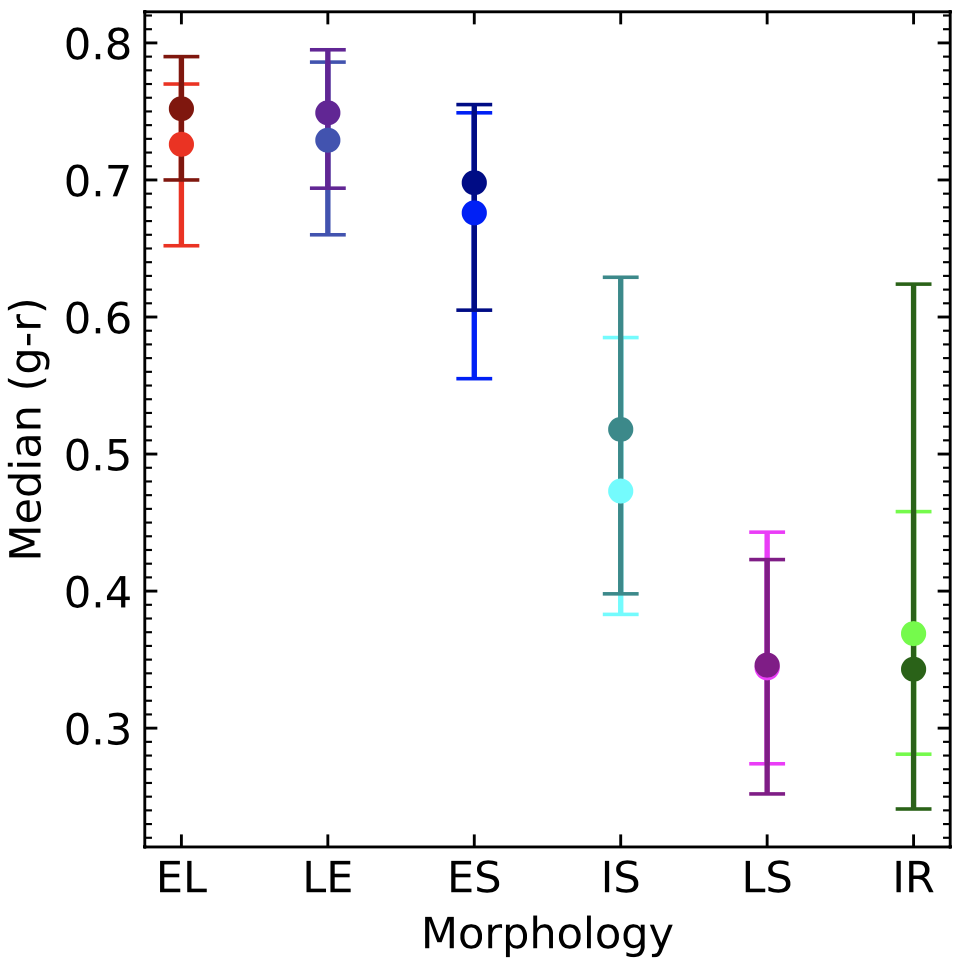}
  		}
  		\hspace{0.3cm}
  		\subfigure{
  			\includegraphics[width=0.3\linewidth]{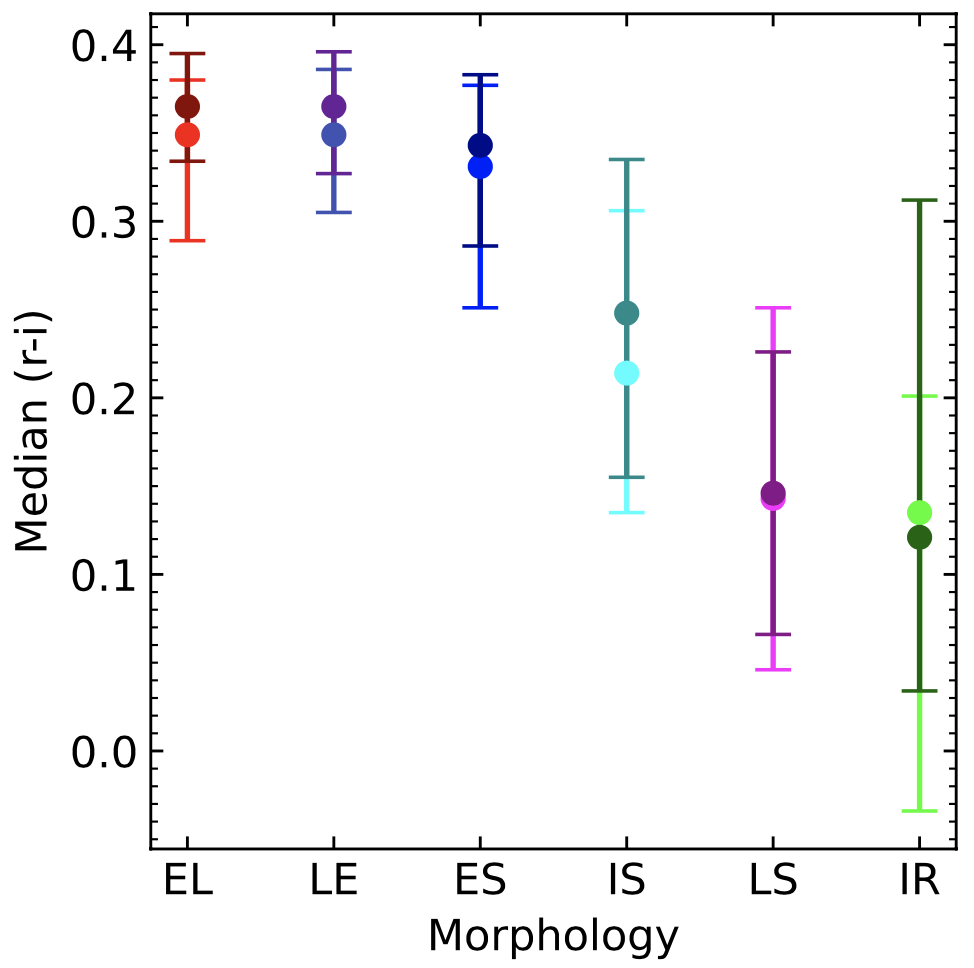}
  		}
  		\hspace{0.3cm}
  		\subfigure{
  			\includegraphics[width=0.3\linewidth]{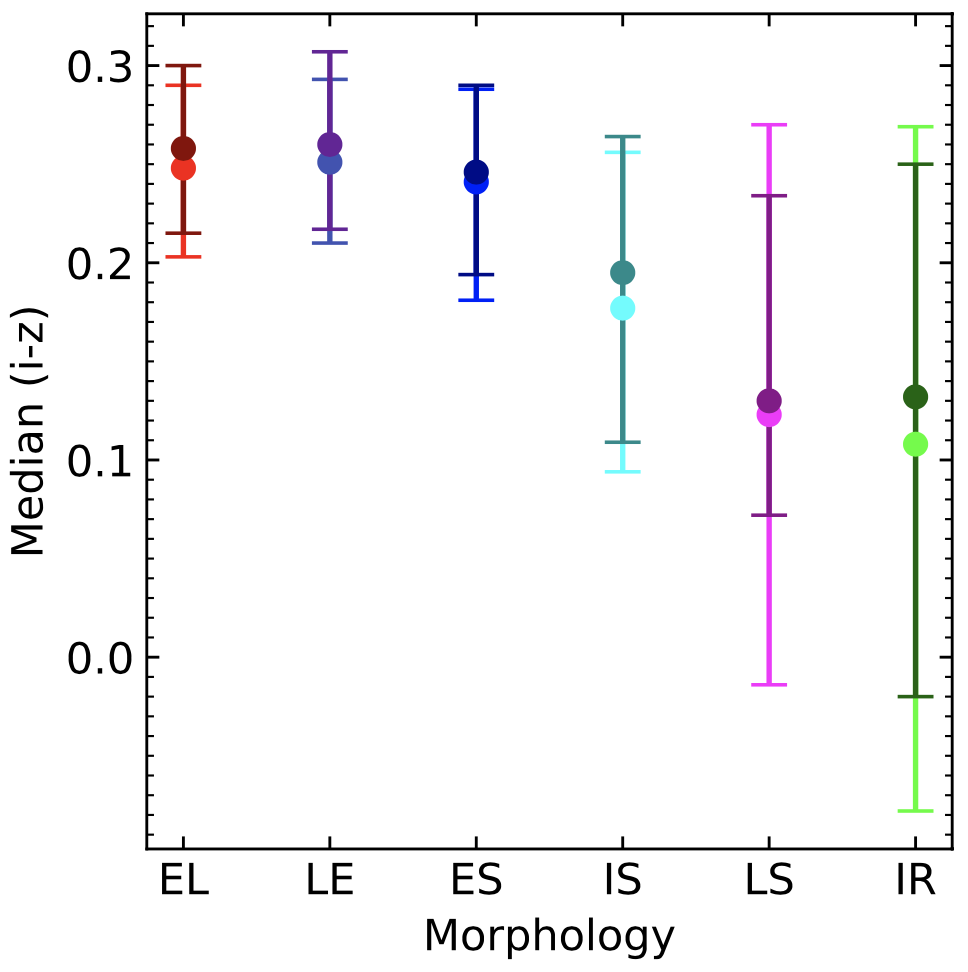}
  		}
  			\hspace{-0.2cm}
\caption{Variation of median colours with morphology for field (light 
colour) and group (dark colour) galaxies. The bars in each measurement 
indicate the 16th and 84th percentiles of the estimate.}
\label{M}
\end{figure}
 
\begin{table}[h!]
\centering
\caption{Median colours for Field (F) and Group (G) types of Elliptical (EL), 
Lenticular (LE), Early-type spiral (ES), Intermediate-type spiral (IS), 
Late-type spiral (LS) and Irregular (IR) galaxies. The associated 
uncertainties are the 16th and 84th percentiles of the estimates.}
\setlength{\tabcolsep}{0.3pc}
  		\begin{center}  
  			\scalebox{0.9}{
  				\begin{tabular}{ccccccccccccc}  
  					\toprule
  					\toprule
  					&  \multicolumn{2}{c} {EL} &  \multicolumn{2}{c}{LE} & 
  					\multicolumn{2}{c}{ES} & \multicolumn{2}{c}{IS} & \multicolumn{2}{c}{LS}& \multicolumn{2}{c}{IR}\\
  					\cmidrule(lr){2-3} \cmidrule(lr){4-5} \cmidrule(lr){6-7} \cmidrule(lr){8-9} \cmidrule(lr){10-11} \cmidrule(lr){12-13}
  					Colour & F & G & F & G & F & G & F & G& F & G& F & G \\
  					{(1)} & (2) & {(3)} & {(4)} & {(5)} & {(6)} & {(7)} & {(8)} & {(9)}& {(10)}& {(11)}& {(12)}& {(13)}\\
  					\midrule
  					$B-V$ & $0.62^{+0.06}_{-0.09}$ & $0.65^{+0.06}_{-0.07}$ & $0.63^{+0.06}_{-0.08}$ & $0.65^{+0.06}_{-0.07}$ & 
  					$0.61^{+0.08}_{-0.14}$ & $0.64^{+0.06}_{-0.11}$ & $0.41^{+0.14}_{-0.10}$&$0.46^{+0.15}_{-0.15}$&
  					$0.30^{+0.09}_{-0.09}$&$0.28^{+0.12}_{-0.10}$&$0.29^{+0.14}_{-0.12}$& $0.23^{+0.33}_{-0.11}$ \\[4pt]
  					$B-R$ & $0.96^{+0.07}_{-0.11}$ & $0.99^{+0.07}_{-0.08}$& $0.97^{+0.09}_{-0.10}$& $1.00^{+0.08}_{-0.09}$& 
  					$0.95^{+0.11}_{-0.19}$&$0.98^{+0.09}_{-0.13}$&$0.70^{+0.20}_{-0.14}$&$0.78^{+0.21}_{-0.20}$&
  					$0.54^{+0.12}_{-0.11}$&$0.52^{+0.15}_{-0.12}$&$0.54^{+0.12}_{-0.17}$&$0.44^{+0.44}_{-0.12}$\\[4pt]
  					$u-g$ & $1.56^{+0.18}_{-0.22}$ & $1.63^{+0.15}_{-0.19}$ & $1.59^{+0.18}_{-0.20}$ & $1.64^{+0.16}_{-0.17}$ & 
  					$1.38^{+0.24}_{-0.27}$&$1.46^{+0.21}_{-0.24}$&$1.00^{+0.20}_{-0.15}$&$1.06^{+0.20}_{-0.19}$&
  					$0.75^{+0.23}_{-0.11}$&$0.81^{+0.13}_{-0.12}$&$0.83^{+0.30}_{-0.12}$&$0.86^{+0.48}_{-0.18}$\\[4pt]
  					$g-r$ & $0.73^{+0.04}_{-0.07}$& $0.75^{+0.04}_{-0.05}$ & $0.73^{+0.06}_{-0.07}$& $0.75^{+0.05}_{-0.05}$ & 
  					$0.68^{+0.07}_{-0.12}$&$0.70^{+0.06}_{-0.09}$&$0.47^{+0.11}_{-0.09}$&$0.52^{+0.11}_{-0.12}$&
  					$0.34^{+0.10}_{-0.07}$&$0.35^{+0.08}_{-0.09}$&$0.37^{+0.09}_{-0.09}$&$0.34^{+0.28}_{-0.10}$\\[4pt]
  					$r-i$ & $0.35^{+0.03}_{-0.06}$& $0.37^{+0.03}_{-0.03}$ &$0.35^{+0.04}_{-0.04}$& $0.37^{+0.03}_{-0.04}$ & 
  					$0.33^{+0.05}_{-0.08}$&$0.34^{+0.04}_{-0.06}$&$0.21^{+0.09}_{-0.08}$&$0.25^{+0.99}_{-0.99}$&
  					$0.14^{+0.11}_{-0.10}$&$0.15^{+0.08}_{-0.08}$&$0.13^{+0.07}_{-0.17}$&$0.12^{+0.19}_{-0.09}$\\[4pt]
  					$i-z$ & $0.25^{+0.04}_{-0.04}$ & $0.26^{+0.04}_{-0.04}$& $0.24^{+0.04}_{-0.04}$& $0.25^{+0.05}_{-0.04}$ & 
  					$0.24^{+0.05}_{-0.06}$&$0.25^{+0.04}_{-0.05}$&$0.18^{+0.08}_{-0.08}$&$0.19^{+0.07}_{-0.09}$&
  					$0.12^{+0.15}_{-0.14}$&$0.13^{+0.10}_{-0.06}$&$0.11^{+0.16}_{-0.19}$&$0.13^{+0.12}_{-0.15}$\\
  					\bottomrule
  				\end{tabular}
  			}
\end{center}
\label{mc}  
\end{table}

  \begin{table}[h!]
 	\centering
 	\caption{Number and percentage variations in colour-magnitude planes 
for field (F) and group (G) galaxies.}
 	\setlength{\tabcolsep}{0.2pc}
 	\begin{center}  
 		\scalebox{0.7}{
 			\begin{tabular}{ccccccccccccc}  
 				\toprule
 				\toprule
 				&  \multicolumn{2}{c} {EL} &  \multicolumn{2}{c}{LE} & 
 				\multicolumn{2}{c}{ES} & \multicolumn{2}{c}{IS} & \multicolumn{2}{c}{LS}& \multicolumn{2}{c}{IR}\\
 				\cmidrule(lr){2-3} \cmidrule(lr){4-5} \cmidrule(lr){6-7} \cmidrule(lr){8-9} \cmidrule(lr){10-11} \cmidrule(lr){12-13}
 				Position & F & G & F & G & F & G & F & G& F & G& F & G \\
 				{(1)} & (2) & {(3)} & {(4)} & {(5)} & {(6)} & {(7)} & {(8)} & {(9)}& {(10)}& {(11)}& {(12)}& {(13)}\\
 				\midrule
 				Red sequence & $2 (0.81\%)$ & $13 (1.91\%)$ & $22 (3.94\%)$ & $54 (4.29\%)$ & 
 				$30 (1.90\%)$ & $48 (2.16\%)$ & $5 (0.30\%)$&$7 (0.61\%)$&
 				$2 (1.82\%)$&$2 (2.67\%)$&$1 (2.63\%)$& $3 (6.00\%)$ \\[3pt]
 				Blue cloud& $59 (23.89\%)$ & $83 (12.22\%)$ & $91 (16.31\%)$ & $113 (8.97\%)$ & 
 				$747 (47.37\%)$&$747 (33.63\%)$&$1563 (92.76\%)$&$1025 (88.90\%)$&
 				$107 (97.27\%)$&$71 (94.67\%)$&$35 (92.11\%)$&$44 (88.00\%)$\\[3pt]
 				Green valley & $186 (75.30\%)$ & $583 (85.86\%)$& $445 (79.75\%)$& $1092 (86.74\%)$& 
 				$800 (50.73\%)$&$1426 (64.21\%)$&$117 (6.94\%)$&$121 (10.49\%)$&
 				$1 (0.91\%)$&$2 (2.67\%)$&$2 (5.26\%)$&$3 (6.00\%)$\\[3pt]
 				Total & $247 (100\%)$& $679 (100\%)$ & $558 (100\%)$& $1259 (100\%)$ & 
 				$1577 (100\%)$&$2221 (100\%)$&$1685 (100\%)$&$1153 (100\%)$&
 				$110 (100\%)$&$75 (100\%)$&$38 (100\%)$&$50 (100\%)$\\
 				\bottomrule
 			\end{tabular}
 		}
\end{center}
\label{GV}  
 \end{table}

\section{Discussion}
\label{secIV}
From Section \ref{gm} it is observed that the Early-type spiral galaxies 
dominate the galaxy population ($\sim 39\%$) followed by the intermediate-type 
spiral ($\sim 29\%$), Lenticular ($\sim 19\%$), Elliptical ($\sim 10\%$), 
Late-type spiral ($2\%$) and Irregular ($1\%$) galaxies. This supports the 
results of Ref.\ \cite{vazquez2022sdss} in which it was obtained that spiral 
galaxies dominate the galaxy population using a sample of $4\,614$ galaxies 
from the MaNGA survey. Similarly, this is in line with the studies by 
Refs.\ \cite{fischer2019sdss,sanchez2021sdss} presenting the MaNGA 
PyMorph photometric Value Added Catalogue (MPP-VAC-DR17) and the MaNGA Deep 
Learning Morphological VAC (MDLM-VAC-DR17) as detailed in 
Ref.\ \cite{sanchez2021sdss}, as they obtained that Late-type galaxies 
(in this case ES, IS, LS, and IR) dominate the galaxies population than the 
Early-type galaxies (in this case EL and LE). 
  
From Section \ref{en}, it is well established that $\sim73\%$, 
$\sim69\%$, $\sim58\%$, and $\sim57\%$ of EL, LE, ES and IR are in a group 
environment while $\sim 59\%$ and $\sim 60\%$ of IS and LS reside in the 
field environment. This result is in line with the findings by Refs.\
\cite{dressler1980galaxy,park2007environmental}, in which one side of the 
existence of a relation between environment and galaxy morphology was 
established when studying galaxies in 55 rich clusters. They further obtained 
that the elliptical galaxies tend to live in a dense environment while spirals 
in a less dense environment. Similarly, the majority of studies observed a 
varying fractions of spiral and elliptical galaxies in different environment 
obtaining $>50\%$ of elliptical galaxies to be in group environment when 
compared to $<40\%$ of spiral galaxies, again $>50\%$ of spirals to be in 
the field environment while less than $40\%$ for the case of elliptical 
galaxies \cite{cluver2020galaxy,nantais2013morphology,wilman2012relation,
van2007evolution,pearson2021galaxy,van2000hubble,bouwens2007uv}. 
We observe that not all spiral galaxies exist preferentially in the field environment 
but the Intermediate-type and Late-types spirals while Early-type spirals exist 
preferentially in the group environment.
  
The observations from KS and AD statistical tests in Tables \ref{KS} and 
\ref{AD} respectively indicate that the galaxies' colours have the average KS 
statistics (p-value) of \num{0.20} (\num{2.60e-03}), \num{0.16} 
(\num{5.75e-05}), \num{0.12} (\num{1.44e-04}), \num{0.15} (\num{12.25e-08}), 
\num{0.16} (\num{0.31}) and \num{0.21} (\num{0.36}) for EL, LE, ES, IS, LS and 
IR respectively as well as average AD statistics of \num{21.86} 
(\num{5.05e-05}), \num{25.04} (\num{1.00e-05}), \num{43.49} (\num{1.00e-05}), 
\num{39.01} (\num{1.00e-08}), \num{1.13} (\num{0.162}) and \num{0.97} 
(\num{0.155}) for EL, LE, ES, IS, LS and IR respectively. Since the average 
KS and AD statistics are much greater than zero and the p-values are less 
than \num{0.05}, this implies that the colours of EL, LE, ES and IS comes 
from different populations. In contrast, as the average KS and AD statistics 
are closer to zero and the p-values are greater than \num{0.05} for colours 
of  LS and IR, the differences are not statistically significant and 
the null hypothesis of these populations coming from the same parent 
population cannot be rejected. Hence there 
is a weak dependence of LS and IR's colours on the environment. 
This is in consistence with the observations from the distributions in 
Figs.~\ref{BV}, \ref{BR}, \ref{ug}, \ref{gr}, \ref{ri}, and \ref{iz}.
  
From Table \ref{mc} and Fig.~\ref{M}, it is well observed that group EL, LE, 
IS and SE are redder than the field ones. This implies that group early types 
are redder than field galaxies, supporting the results from 
Refs.\ \cite{girardi2003galaxies,balogh2009colour,pandey2020exploring,
cramer2024resolved,coenda2020effects}. Ref.\ \cite{girardi2003galaxies} 
pointed out that early-type galaxies in groups are redder than field ones 
when analysing the galaxy sample from Nearby Optical Galaxy (NOG) catalogue 
\cite{giuricin2000nearby}. For the case of LS and IR there is no clear trend. 
This observation is made from the associated overlapping of uncertainty 
which does not make any significant difference. These results imply that the
dependence of galaxies' colours on the environment is influenced by morphology.
  
Observations from the colour-magnitude planes in Fig.\ \ref{CM} and 
Table~\ref{GV} are that the EL, LE and ES galaxies populate the green valley 
while the IS, LS and IR populate the blue sequence supporting the results by 
Refs.\ \cite{sampaio2022blue,eales2018causes,schawinski2014green}, that 
morphological changes is associated with colour transformations. It is 
observed that group galaxies are more found in green valley and red sequence 
than field galaxies which are more found in blue cloud. Using the chi-square 
statistical test ($\chi^2$) the following $\chi^2$ statistics and p-values 
were obtained: \num{19.86} (\num{4.87e-05}), \num{14.52} (\num{7.05e-04}), 
\num{73.10} (\num{1.34e-16}), \num{12.97} (\num{1.53e-03}), \num{1.03} 
(\num{0.598}), \num{0.60} (\num{0.741}) for EL, LE, ES, IS, LS and IR 
respectively. Since the p-values are less than the standard p-value in 
statistics ($0.05$) for EL, LE, ES and IS the difference in number of galaxies 
between field and group is significant. On the other hand, for the case of LS 
and IR the difference is insignificant since the p-values are greater than 
$0.05$. It is obvious that the blue cloud to red sequence transformation 
depends on the environment for the Elliptical, Lenticular, Early-type spiral 
and Intermediate-type while for Late-type and Irregular galaxies the 
transformation is not influenced by the environment. The study revealed that 
the dependence of the colour-magnitude planes on the environment is influenced 
by morphology.
 
\section{Summary and Conclusion}
  \label{secV}
In this study, we used integral field spectroscopy (IFS) data from Mapping 
Nearby Galaxies at Apache Point Observatory (MaNGA) to investigate if 
morphology influences the environmental dependence of galaxy colours and the 
colour-magnitude planes. The galaxies were classified into six morphologies 
where the images are shown in Fig.~\ref{IM} at which cD and E (Elliptical), 
SO (Lenticular), Sa, Sab, and Sb (Early-type spiral), Sbc and Sc 
(Intermediate-type spiral), Scd and Sd (Late-type spirals), Sdm, Sm and Irr 
(Irregular) detailed in Refs.\ \cite{vazquez2022sdss,sanchez2022sdss} which 
follow Ref.~\cite{hubble1926} classification scheme. The galaxy environments
are quantified for all six morphologies that are Elliptical (EL), Lenticular 
(LE), Early-type spirals (ES), Intermediate-type spirals (IS), Late-type 
spirals (LS) and Irregular (IR) using the Galaxy Environment for MaNGA Value 
Added Catalogue (GEMA-VAC) as detailed in Refs.\ \cite{argudo2015catalogues,
etherington2015measuring,wang2016elucid} where galaxies having no nearby 
galaxy (GS $=1$) are field and the ones with more than one 
neighbour (GS $\geq2$) are group galaxies. The following numbers of galaxies 
were obtained, EL: $247\,(26.67\%)$, $679\,(73.33\%)$,  LE: $558\,(30.71\%)$, 
$1259\,(69.29\%)$, ES: $1577\,(41.52\%)$, $2221\,(58.48\%)$, 
IS: $1685\,(59.37\%)$, $1153\,(40.63\%)$, LS: $110\,(59.46\%)$, 
$75\,(40.54\%)$, IR: $38\,(43.18\%)$, $50\,(56.82\%)$ for field, group 
galaxies. These numbers were used to compare the distributions between field 
(F) and group (G) environments for six colour indices ($B-V$, $B-R$, $u-g$, 
$g-r$, $r-i$, $i-z$) as shown by Figs.~\ref{BV}, \ref{BR}, \ref{ug}, 
\ref{gr}, \ref{ri} and \ref{iz}. We employed the Kolmogorov-Smirnov 
(KS) \cite{hodges1958significance,harari2009kolmogorov} and Anderson–Darling 
(AD) \cite{anderson1952asymptotic,pettitt1976two,scholz1987k,
babu2006astrostatistics} statistical tests to assess of the degree of 
similarities or differences between field and group environments, shown by 
Tables \ref{KS} and \ref{AD} respectively and the median colours were compared 
in Fig.~\ref{M} and Table \ref{mc}. Further, the colour-magnitude planes in 
Fig.~\ref{CM} were analysed, in which the red sequence and blue cloud were 
defined based on Refs.\ \cite{ blanton2005properties,dhiwar2023witnessing,
bom2024extended} criteria given by Eqs.~\eqref{CM1} and \eqref{CM2}. 
Table \ref{GV} shows the positioning of galaxies on the colour-magnitude 
diagram in which we employed the $\chi^2$ statistical test to assess the 
significance of the difference obtained between field and group environment. 
Together with the already established results this study revealed the 
following:
\begin{itemize}
    	\item Not all spiral galaxies exist preferentially in field 
environments but the Intermediate-type and Late-type spirals while 
Early-type exist preferentially in group environments.
    	\item The galaxy colours depend on the environment when Elliptical, 
Lenticular, Early-type and Intermediate-type spirals are considered, but for 
Late-type spiral and Irregular galaxies, the dependence of colour on the 
environment is very weak.
    	\item The blue cloud to red sequence transformation of Elliptical, 
Lenticular, Early-type and Intermediate-type spirals depends on the environment 
while for Late-type and Irregular galaxies, there is a weak dependence.
\end{itemize}
This study revealed that the dependence of colours and colour-magnitude planes 
on the environment is influenced by morphology. In the future, the study will 
be extended to high redshift to analyse if the redshift influences the observed 
relations.

\section*{Acknowledgements} PP acknowledges support from The Government of 
Tanzania through the India Embassy, Mbeya University of Science and Technology 
(MUST) for Funding. UDG is thankful to the Inter-University Centre for 
Astronomy and Astrophysics (IUCAA), Pune, India for the Visiting 
Associateship of the institute. Funding for the Sloan Digital Sky Survey IV 
has been provided by the Alfred P.~Sloan Foundation, the U.S.~Department of 
Energy Office of Science, and the Participating Institutions. SDSS-IV 
acknowledges support and resources from the Center for High-Performance 
Computing at the University of Utah. SDSS-IV is managed by the Astrophysical 
Research Consortium for the Participating Institutions of the SDSS 
Collaboration, including the Brazilian Participation Group, the Carnegie 
Institution for Science, Carnegie Mellon University, Center for Astrophysics -- 
Harvard \& Smithsonian, the Chilean Participation Group, the French 
Participation Group, Instituto de Astrofísica de Canarias, The Johns Hopkins 
University, Kavli Institute for the Physics and Mathematics of the Universe 
(IPMU)/University of Tokyo, the Korean Participation Group, Lawrence Berkeley 
National Laboratory, Leibniz Institut f\"ur Astrophysik Potsdam (AIP), 
Max-Planck-Institut f\"ur Astronomie (MPIA Heidelberg), Max-Planck-Institut 
f\"ur Astrophysik (MPA Garching), Max- Planck-Institut f\"ur 
Extraterrestrische Physik (MPE), National Astronomical Observatories of China,
New Mexico State University, New York University, University of Notre Dame, 
Observatário Nacional/MCTI, The Ohio State University, Pennsylvania State 
University, Shanghai Astronomical Observatory, United Kingdom Participation 
Group, Universidad Nacional Autónoma de México, University of Arizona, 
University of Colorado Boulder, University of Oxford, University of 
Portsmouth, University of Utah, University of Virginia, University of 
Washington, University of Wisconsin, Vanderbilt University, and Yale 
University.

\end{document}